\documentclass[11pt]{article}
\usepackage{jheppub}
\usepackage{amsmath}
\usepackage{amssymb}
\usepackage{braket}
\usepackage{bbold}
\usepackage{pifont}
\usepackage{graphicx} 
\usepackage{bmpsize}

\title{Structures in multiple spin-2 interactions}
\author{Oliver Baldacchino~\&}
\author{Angnis~Schmidt-May}
\affiliation{Institut f\"ur Theoretische Physik, Eidgen\"ossische Technische Hochschule Z\"urich,\\
Wolfgang-Pauli-Strasse 27, 8093 Z\"urich, Switzerland}   
\emailAdd{oliverb@student.ethz.ch}    
\emailAdd{angniss@itp.phys.ethz.ch}

\abstract{
We study generalisations of ghost-free bimetric theory which involve more than two spin-2 fields. The consistent interactions can enter in the form of two different couplings and in the majority of this work we concentrate on the simpler one. The corresponding action involves one metric coupled to $N$ tensor fields which do not interact with each other.
We derive maximally symmetric solutions to the multimetric equations of motion and identify the mass eigenstates in the linearised theory around these backgrounds. Our results are then applied to the problem of singling out multimetric models which possess certain additional structures. In particular, we look for a relation between scale invariant background solutions, the perturbative emergence of Weyl invariance and the presence of partially massless spin-2 fields in the linear theory.
Our findings generalise known results in bimetric theory and allow us to point out similarities and differences between the bi- and multimetric models.
}


\newcommand{\beqn}{\begin{eqnarray}}
\newcommand{\eeqn}{\end{eqnarray}}
\newcommand{\dd}{\mathrm{d}}
\newcommand{\nn}{\nonumber}
\newcommand{\Tr}{\mathrm{Tr}}
\newcommand{\gmn}{g_{\mu\nu}}
\newcommand{\fmn}{f_{\mu\nu}}
\newcommand{\hmn}{h_{\mu\nu}}

\begin{document} 
\maketitle
\flushbottom

\section{Introduction}
\subsection{Background and motivation}

General relativity is the nonlinear field theory for a self-interacting massless spin-2 field. For fields with spin $0$, $1/2$ and $1$, which make up the Standard Model of Particle Physics, it has long been known how to construct mass terms and how to couple the fields to the massless graviton. The corresponding models are all realised in nature. 
In the spin-2 case it is much more difficult to write down a mass term and interactions with gravity. The reason for this complication is that generic descriptions of massive spin-2 field, like many other modifications of general relativity, are inconsistent in the sense that they contain unwanted additional degrees of freedom (``ghosts") leading to instabilities.

In 1939, Fierz and Pauli succeeded in constructing the linear (i.e.~non-interacting) theory for massive spin-2 fields in flat space~\cite{Fierz:1939ix}. Unfortunately, all attempts to extend this description by nonlinear interactions for the field failed and in 1972 Boulware and Deser presented a theorem stating that any interacting theory for massive spin-2 fields would suffer from a fatal ghost instability~\cite{Boulware:1973my}. This result was essentially trusted until 2010, when de Rham, Gabadadze and Tolley discovered a loophole in the assumptions entering the proof by Boulware and Deser. Based on the preceding analyses of Ref.~\cite{Creminelli:2005qk, ArkaniHamed:2002sp}, they proposed a nonlinear theory for a self-interacting massive spin-2 field~\cite{deRham:2010ik, deRham:2010kj}. Shortly after, it was confirmed that this theory for massive gravity indeed avoids the problematic ghost instability~\cite{Hassan:2011hr, Hassan:2011tf, Hassan:2011ea, Hassan:2012qv}.
Since then, nonlinear massive gravity has been argued to suffer from problems related to acausality and superluminality~\cite{Deser:2013eua, Deser:2015wta}.
Moreover, the ghost-free action involves a non-dynamical reference metric in addition to the massive spin-2 degrees of freedom, which seems somewhat unusual from a field theoretical perspective. 
Trying to give a meaning to the reference metric, Hassan and Rosen demonstrated that it can be given its own dynamics without reintroducing the Boulware-Deser ghost, which resulted in the first consistent bimetric theory~\cite{Hassan:2011zd}. This fully dynamical theory for two tensor fields describes the nonlinear interactions of a massive and a massless spin-2 field. 
For a certain class of solutions the zero-mass limit in massive gravity is discontinuous~\cite{vanDam:1970vg, Zakharov:1970cc}. This discontinuity can be avoided in certain energy regimes by the means of nonlinear effects \cite{Vainshtein:1972sx} and, as has recently been shown, it is absent on other types of solutions~\cite{deRham:2016plk}.
On the other hand, bimetric theory possesses a smooth general relativity limit with universal validity~\cite{Baccetti:2012bk, Hassan:2014vja}. 
Earlier work on bimetric theories includes Ref.~\cite{Rosen:1975kk, Aragone:1971kh, Isham:1971gm, Damour:2002ws, Buchbinder:1999ar, Chamseddine:1978yu, Aragone:1979bm}.

Due to these achievements in recent years, our understanding of field theories for massive spin-2 fields has significantly improved and the ghost-free theories have been the subject of many investigations, including applications to cosmology. 
For an exhaustive list of references, we refer the reader to the review articles~\cite{Hinterbichler:2011tt, deRham:2014zqa, Schmidt-May:2015vnx}.

One of the most obvious continuative questions is whether one can also construct theories involving more than two tensor fields. 
It is an established result that consistent theories with more than one interacting massless spin-2 field cannot exist\footnote{Note however that in three spacetime dimensions a theory of interacting massless spin-2 fields has been constructed in the Chern-Simons formulation of gravity~\cite{Gwak:2015vfb}.}\cite{Boulanger:2000rq}, but no such strong statement has been made for massive fields. Hinterbichler and Rosen were the first to study generalisations of ghost-free bimetric theory involving more than one massive spin-2 field~\cite{Hinterbichler:2012cn}. They worked in the vierbein formulation and proposed generalised vertices for multiple vierbein interactions which generically could not be rewritten in terms of metric variables (see also Ref.~\cite{Hassan:2012wt, Nomura:2012xr, Deffayet:2012zc}). It later turned out that in general such couplings suffer from the Boulware-Deser ghost~\cite{deRham:2015cha}, while some particular interaction terms considered in Ref.~\cite{Hinterbichler:2012cn} still avoid the instability. The consistent models possess a formulation in terms of metrics and are essentially built from multiple copies of the ghost-free bimetric action. The interactions come in two different structures which we will refer to as ``chain" and ``centre" couplings throughout this paper. In particular, couplings of several metrics in a loop are forbidden.\footnote{In fact, it was already noticed in Ref.~\cite{Hinterbichler:2012cn} and also in \cite{Hassan:2012wt, Nomura:2012xr, Scargill:2014wya} that the loop couplings are not consistent.}
The consistent models have recently been studied by Ferreira, Noller and Scargill~\cite{Noller:2013yja, Noller:2015eda, Scargill:2015wxs} whose main focus lied on the so-called decoupling limit and the identification of energy scales relevant for the effective field theory picture. Moreover, the cosmology of three interacting metrics was investigated in~\cite{Luben:2016lku}.

Nonlinear massive gravity, bimetric and multimetric theory all possess an intrinsic strong-coupling scale and should be interpreted as effective theories valid below this scale (see, for instance,~\cite{Scargill:2015wxs}). It is therefore worth mentioning that, in principle, there exists a larger class of spin-2 theories whose low-energy regimes are free from ghost instabilities. Here we choose to focus on actions that do not propagate ghosts on any scale at the classical level and this is what we refer to when we use the phrases ``consistent" or ``ghost-free".

In this work we shall revisit the two classes of ghost-free multimetric couplings and generalise several results obtained in bimetric theory to models with more than one massive spin-2 field.  
We begin by giving a very brief and non-technical overview of Hassan and Rosen's bimetric theory, focussing on properties that we intend to recover in multimetric interactions.

\subsection{Short review of bimetric theory}

The classically consistent bimetric action, containing Einstein-Hilbert terms and interactions for two tensor fields $\gmn$ and $\fmn$, is given by,
\beqn\label{actbmt}
S[g,f]=m_g^2\int\dd^4x\Big( \sqrt{g}~R(g)+\alpha^2 \sqrt{f}~R(f)-2m^2V(g,f;\beta_n)\Big)\,,
\eeqn
in which $m_g^2$ and $\alpha^2m_g^2$ are the respective Planck masses and $m$ sets the mass scale of interactions. The potential is required to have a specific structure in order to avoid the Boulware-Deser ghost instability and contains five free parameters $\beta_n$. Its explicit form will be provided in the context of multimetric theory later on. 

The equations of motion obtained from the bimetric action are of the form,
\begin{subequations}\label{eombt}
\beqn
\mathcal{G}_{\mu\nu}(g)+m^2 V_{\mu\nu}(g,f;\beta_n)=0\,,\qquad
\mathcal{G}_{\mu\nu}(f)+\frac{m^2}{\alpha^2}V_{\mu\nu}(f, g;\beta_{4-n})=0\,,
\eeqn
\end{subequations} 
where $\mathcal{G}_{\mu\nu}(g)=R_{\mu\nu}(g)-\frac{1}{2}\gmn R(g)$ is the Einstein tensor, obtained from varying the kinetic terms, and $V_{\mu\nu}$ denotes the contributions arising from the interaction potential. 
The simplest solutions to these equations are the proportional backgrounds, $\bar{f}_{\mu\nu}=c^2\bar{g}_{\mu\nu}$ with constant $c$. They correspond to maximally symmetric Einstein metrics $\bar{g}_{\mu\nu}$ whose curvatures satisfy $R_{\mu\nu}(\bar{g})=\Lambda\bar{g}_{\mu\nu}$. Around these backgrounds it is straightforward to compute the mass spectrum of bimetric theory. One decomposes the metrics  into background and small fluctuations, $\gmn=\bar{g}_{\mu\nu}+\delta\gmn$ and $\fmn=c^2\bar{g}_{\mu\nu}+\delta\fmn$, and linearises the equations of motion in $\delta\gmn$ and~$\delta\fmn$. The resulting expressions can be diagonalised into two sets of decoupled equations of the form~\cite{Hassan:2012wr},
\beqn\label{linbimeq}
\bar{\mathcal{E}}^{~~\rho\sigma}_{\mu\nu}\delta G_{\rho\sigma}=0\,,\qquad
\bar{\mathcal{E}}^{~~\rho\sigma}_{\mu\nu}\delta  M_{\rho\sigma} - \tfrac{m_\mathrm{FP}^2}{2}\big( \delta M_{\mu\nu}-\delta M\bar{g}_{\mu\nu}\big)=0\,,
\eeqn
where $\delta G_{\mu\nu}\equiv\delta\gmn+\alpha^2\delta\fmn$ and $\delta M_{\mu\nu}\equiv\delta\fmn-c^2\delta\gmn$. The two-derivative operator~$\bar{\mathcal{E}}^{~~\rho\sigma}_{\mu\nu}$, whose precise form will be given later, arises from linearising the Einstein tensor and cosmological constant contributions. The contributions proportional to $m_\mathrm{FP}^2$ come from the interaction potential and vanish for $m=0$. 
The two sets of equations in~(\ref{linbimeq}) describe a massless and a massive spin-2 field, respectively, propagating on maximally symmetric backgrounds $\bar{g}_{\mu\nu}$. The Fierz-Pauli mass $m_\mathrm{FP}$ and the background curvature $\Lambda$ are both functions of the parameters in the bimetric action. 

The symmetries of the linear spin-2 equations above are well-known. The massless equations are invariant under linearised diffeomorphisms, $\Delta(\delta G_{\mu\nu})=\bar{\nabla}_\mu\xi_\nu+\bar{\nabla}_\nu\xi_\mu$, where $\bar{\nabla}$ is the covariant derivative compatible with the background metric $\bar{g}_{\mu\nu}$. This gauge symmetry reduces the number of physical degrees of freedom in $\delta G_{\mu\nu}$ to two.
Generically, the massive equations do not possess any symmetries and propagate the five helicity states of the massive spin-2 field~$\delta M_{\mu\nu}$. 

\paragraph{Additional structure:} Around de Sitter backgrounds with $\Lambda>0$, it is possible for the massive spin-2 equations to exhibit a scalar gauge invariance under particular circumstances.
Namely, if a special relation of mass and cosmological constant is imposed,
\beqn\label{higuchi}
m_\mathrm{FP}^2=\tfrac{2}{3}\Lambda\,,
\eeqn 
the equations for $\delta M_{\mu\nu}$ in~(\ref{linbimeq}) are invariant under the following local transformations with gauge parameter $\phi(x)$~\cite{Deser:1983mm, Deser:2001us},
\beqn\label{PMtrafo}
\Delta(\delta M_{\mu\nu})=\bar{\nabla}_\mu\partial_\nu\phi+\tfrac{\Lambda}{3}\phi\bar{g}_{\mu\nu}\,.
\eeqn
Spin-2 fields whose mass lies on the so-called Higuchi bound~(\ref{higuchi}) have been dubbed ``partially massless" (PM).\footnote{For mass values below the Higuchi bound, $m_\mathrm{FP}^2<\tfrac{2}{3}\Lambda$, the helicity-zero mode of the spin-2 field is a ghost~\cite{Higuchi:1986py}. The partially massless line therefore separates the unitary from the non-unitary region in the $(m_\mathrm{FP}^2,\Lambda)$-plane of the linear spin-2 theory.} The additional gauge symmetry in their equations removes their helicity-zero mode and thereby reduces the number of propagating degrees of freedom to four. For recent work on PM fields see~\cite{Deser:2006zx, Hinterbichler:2014xga, Bonifacio:2015rea, Hinterbichler:2015nua, Cherney:2015jxp}.

An example for a bimetric model whose linear theory contains a PM field has been suggested in Ref.~\cite{Hassan:2012gz} and was further analysed in Ref.~\cite{deRham:2013wv}. 
Besides satisfying the Higuchi relation~(\ref{higuchi}) at the linearised level, this particular model seems to possess a remarkable amount of additional structure, even in its nonlinear interactions. For instance, its proportional background solutions are invariant under constant scale transformations. By this we mean that the proportionality constant $c^2$ in $\bar{f}_{\mu\nu}=c^2\bar{g}_{\mu\nu}$ is not determined by the equations of motion and can therefore be shifted arbitrarily~\cite{Hassan:2012gz}. 
Moreover, as shown in Ref.~\cite{Hassan:2013pca}, the nonlinear equations of motion in this model can be recast as perturbative expansions in curvatures (or derivatives) and the lowest order of these expansions is invariant under local Weyl transformations of the metrics. The symmetry transformations can be extended perturbatively, at least up to sixth order in derivatives~\cite{Hassan:2015tba}. 

The relations between scale invariance of the maximally symmetric solutions, perturbative Weyl invariance and partial masslessness at the linear level are not fully understood. One motivation for our present analysis is to see whether these properties are a coincidence in bimetric theory or whether they can be found also in multimetric interactions, which would suggest the existence of a deeper underlying reason.

\subsection{Outline of strategy and summary of results}

The two consistent types of ghost-free multimetric interactions coincide for models involving up to three fields. In the presence of more fields, we focus on the simpler and more symmetric coupling, in which one central tensor field interacts with $N$ other fields through copies of bimetric interactions. 
For this type of coupling we derive solutions in which all fields are maximally symmetric Einstein metrics and proportional to each other. The mass spectrum of linear fluctuations around these backgrounds is analysed and we calculate the exact solutions for the mass eigenvalues in the case $N=2$.
For general $N$ we obtain expressions for the spin-2 mass eigenstates whose eigenvalues correspond to the roots of a polynomial equation of order $(N+1)$. As expected from the nonlinear Hamiltonian analysis~\cite{Hinterbichler:2012cn}, the linear theory generically contains one massless and $N$ massive spin-2 modes. Moreover, we show that the solutions to the eigenvalue equation greatly simplify when certain symmetry conditions on the interaction parameters are imposed.

We then use these results to identify models with certain additional structures, in analogy to the bimetric case. Focussing on the case $N=2$, we first demand scale invariance of the proportional background solutions. This requirement translates into conditions on the interaction parameters which, in contrast to bimetric theory, do not lead to a unique model but leave us with a subset of free parameters. Our procedure can also be applied to general $N$ and we use it to explicitly derive (highly symmetric) examples of models with scale invariant backgrounds.

Next, starting again from the most general action for three interacting tensor fields, we try to recast the equations in the form of perturbative expansions in curvatures. This turns out to be possible only for a rather restricted set of solutions for which the interactions separate into two bimetric couplings. With this restriction, we derive the expansions and then require their lowest orders to be invariant under local Weyl transformations. Remarkably, the resulting conditions on the interaction parameters are identical to those obtained in demanding scale invariant proportional backgrounds. In the case of general $N$, we show that the most symmetric models derived by requiring scale invariance also lead to expansions whose lowest orders are Weyl invariant.

Finally, we verify that, for the set of models derived via the two independent procedures, one of the linear mass eigenstates possesses an eigenvalue which satisfies~(\ref{higuchi}). It is therefore a partially massless spin-2 field whose equations are invariant under transformations of the form~(\ref{PMtrafo}). As in bimetric theory, imposing the Higuchi bound~(\ref{higuchi}) alone does not uniquely lead to the models with scale invariance and perturbative Weyl symmetry. The remaining linear massive fields in the spectrum violate the Higuchi bound and therefore lead to non-unitarities.

Our findings are visualised in Figure~1 and constitute a direct generalisation of the situation in bimetric theory. The only difference is that the resulting bimetric model is unique, whereas the potential for three metrics still contains free parameters.

   \begin{figure}[h]
    \begin{center}
    \includegraphics[width=250pt]{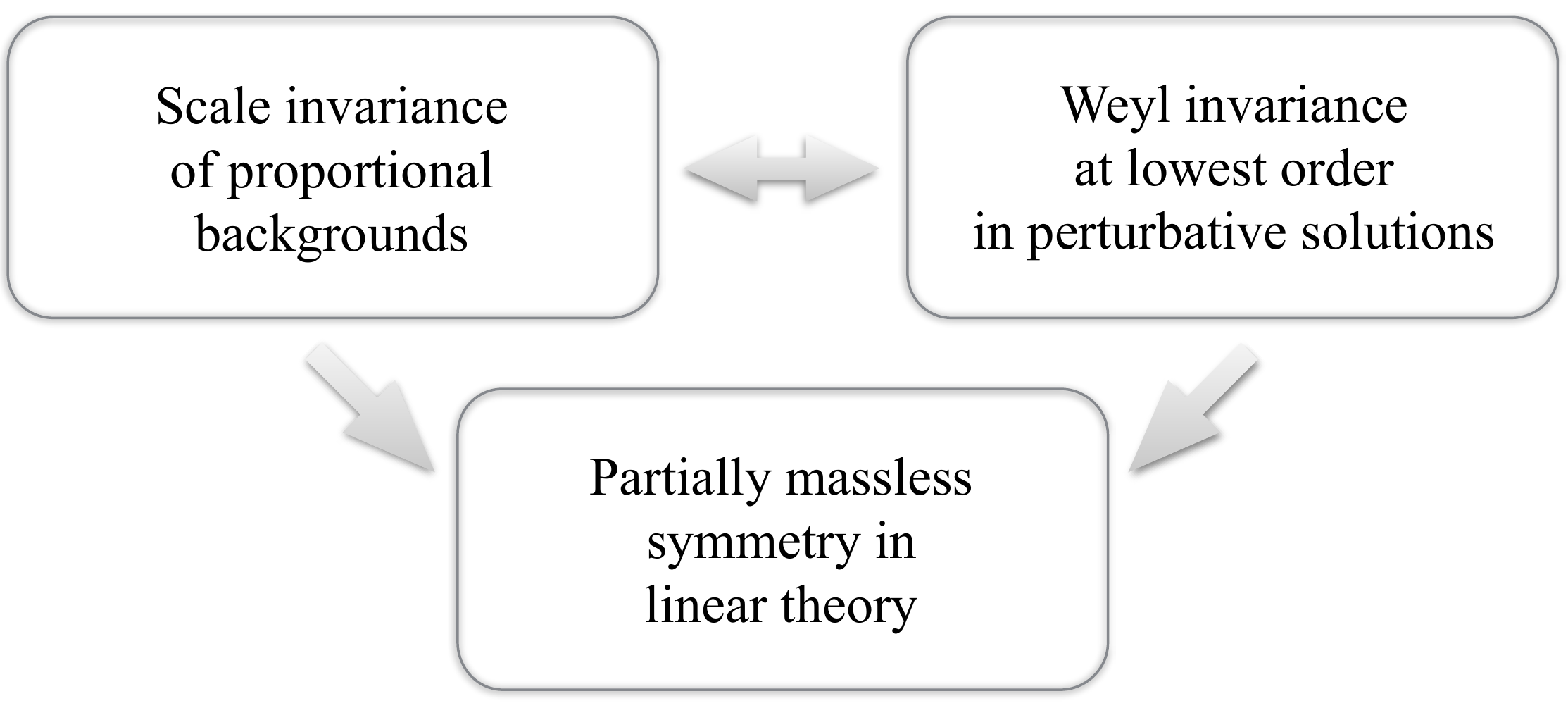}
    \caption{The equivalence of demanding either scale invariance of the proportional backgrounds or perturbative Weyl symmetry. In both cases the mass spectrum contains a PM field. This chart describes the established relations for two and three interacting metrics. For more fields we suspect (but do not conclusively prove) it to look the same.}
    \end{center}
    \end{figure}

Our paper is organised as follows. We begin with a review of consistent multimetric interactions in section~\ref{sec:multi} and from there on focus on the simpler type of coupling. The proportional background solutions and their mass spectrum are derived in section~\ref{sec:pbgms}. In section~\ref{sec:mwas} we apply methods known from the bimetric case to identify multimetric models with the desired additional structures. Our findings are discussed in section~\ref{sec:discussion}.
Appendix~\ref{app:pert} contains technical details in the derivation of the curvature expansions and some properties of the second type of consistent coupling are discussed in appendix~\ref{app:chcoupl}.

\section{Multimetric theory}\label{sec:multi}

Several different analyses have shown that the only ghost-free couplings for more than two tensor fields are simply copies of the consistent bimetric potential. For instance, the results of Ref.~\cite{Hinterbichler:2012cn, deRham:2015cha} obtained for the vierbein formulation of bimetric theory imply the absence of additional couplings for several metrics.\footnote{More precisely, the authors of Ref.~\cite{deRham:2015cha} argued that implementing the so-called symmetric vielbein condition is crucial for the absence of ghost. This observation rules out the more general vielbein couplings suggested in Ref.~\cite{Hinterbichler:2012cn} and leaves only the two types of couplings considered in our present work. We will not repeat the details of the (rather technical) arguments here but refer the reader to the original references.} Analogous arguments given purely in terms of metrics can be found in Ref.~\cite{Hassan:2012wt, Nomura:2012xr, Schmidt-May:2015vnx}. 
As already mentioned in the introduction, ghost-free multimetric interactions can be categorised into two different sets.
In this section we review these two types of couplings for an arbitrary number of interacting tensor fields.

\subsection{Action with centre couplings}

Throughout this paper we will mostly work with the first type of consistent coupling for $(N+1)$ tensor fields which is of the form,
\beqn\label{actcentre}
S_\mathrm{centre}[g,f^i]=m_g^2\int\dd^4x\Bigg( \sqrt{g}~R(g)+\sum_{i=1}^N\alpha^2_i \sqrt{f^i}~R(f^i)
-2m^2\sum_{i=1}^NV(g,f^i;\beta_n^{(i)})\Bigg)\,.
\eeqn
This action contains the spin-2 mass scale $m$ as well as $(N+1)$ Planck masses, parameterised as $\alpha_im_g$ for $i=1,\hdots N$, where $m_g$ is the Planck mass for the metric $\gmn$ which sits in the centre of the coupling.
In this model, the $N$ fields $\fmn^i$ all interact with $\gmn$ but not directly with each other.
The potential therefore consists of multiple copies of bimetric interactions and each term in the sum has the functional form,
\beqn\label{bimpot}
V(g,f;\beta_n)=\sqrt{g}\sum_{n=0}^4\beta_n e_n\big(\sqrt{g^{-1}f}\,\big)\,.
\eeqn
It is precisely this form of interactions that also enters the ghost-free bimetric action~(\ref{actbmt}).
The complete potential in (\ref{actcentre}) contains $5N$ free parameters $\beta_n^{(i)}$ as well as the elementary symmetric polynomials $e_n$ which are scalar-valued functions of a matrix $S$, defined via the recursion relation,
\beqn
e_n(S)=\frac{(-1)^{n+1}}{n}\sum_{k=0}^{n-1}\Tr(S^{n-k})e_k(S)\,,\qquad \text{with}~~e_0(S)=1\,.
\eeqn
Their matrix element is the square-root $S^\mu_{~\nu}=\big(\sqrt{g^{-1}f}\big)^\mu_{~\nu}$ which satisfies $(S^2)^\mu_{~\nu}=g^{\mu\rho}f_{\rho\nu}$. 
Note that, since $e_4(S)=\det S$, the cosmological constant for each $\fmn^i$ is parameterised by $\beta^{(i)}_4$. 
Moreover, since $e_0(S)=1$, the corresponding term for $\gmn$ is given by the sum of all $\beta_0^{(i)}$ and one can set all but one of them to zero without loss of generality. We shall do so in the explicit calculations but, in order to treat the interactions in more symmetric manner, we keep them in most of our general expressions.
The remaining $3N$ parameters $\beta_n^{(i)}$ with $n=1,2,3$ measure true interactions between the two tensor fields.
The potential~(\ref{bimpot}) is invariant under the simultaneous interchange of $\gmn \leftrightarrow\fmn$ and $\beta_n\mapsto\beta_{4-n}$, which implies that in bimetric theory the two tensors are treated on equal footing.
   \begin{figure}[h]
    \begin{center}
    \includegraphics[width=70pt]{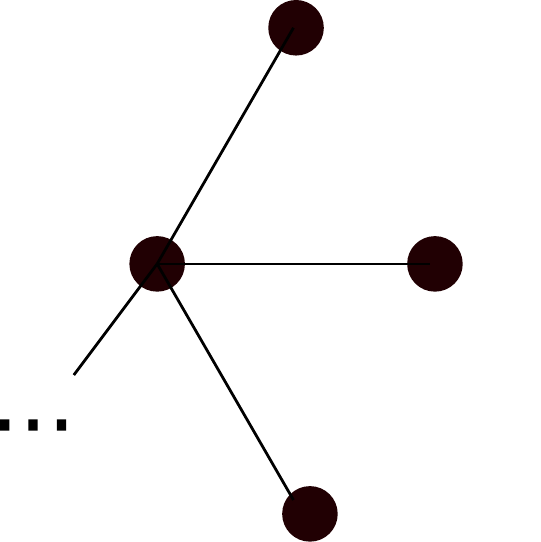}
    \caption{Graph for a centre coupling. The lines represent direct interactions among two tensor fields through the ghost-free potential. The fields themselves are shown as black dots.}
    \end{center}\label{figcentchaint}
    \end{figure}
\\ The centre coupling is visualised in Figure~2.
It is not possible to introduce additional couplings between two of the ``satellite metrics" $\fmn^i$ because such terms would destroy the constraints that remove the Boulware-Deser ghosts~\cite{Hinterbichler:2012cn, deRham:2015cha, Nomura:2012xr}. In general, any type of ``loop coupling" is forbidden in multimetric theory. Nevertheless it is worth mentioning that, before this definite conclusion was reached, some work had been done on cosmology with spin-2 fields coupled in a loop~\cite{Khosravi:2011zi, Tamanini:2013xia}.

The structure of the multimetric action with centre coupling is invariant under the following interchange of any two satellite metrics $\fmn^i$ and $\fmn^j$, along with a simultaneous swap of parameters,
\beqn\label{intsym}
\alpha_i^2\fmn^i\longleftrightarrow\alpha_j^2\fmn^j\,,\qquad
\frac{\beta_n^i}{\alpha^n_i}\longleftrightarrow\frac{\beta_n^j}{\alpha^n_j}\,, \qquad
i,j=1,\hdots N\,.
\eeqn
These $(N-1)!$ invariances are simply a manifestation of the fact that all $\fmn^i$ enter the action in a symmetric way, whereas the central metric $\gmn$ assumes a distinguished position. It is easy to see that couplings in which all $(N+1)$ metrics can be treated symmetrically necessarily introduce loops and are therefore inconsistent. The only exception is the bimetric case with $N=1$. Note, however, that (\ref{intsym}) is not a true discrete symmetry of the action because it also involves changing the parameters. Only models in which the parameters are fixed in a symmetric way show this higher degree of symmetry.

Varying the action~(\ref{actcentre}) with centre couplings with respect to the fields $\gmn$ and $\fmn^i$ results in the following $(N+1)$ sets of equations,
\begin{subequations}\label{eomcenter}
\beqn
\mathcal{G}_{\mu\nu}(g)+m^2\sum_{i=1}^N V_{\mu\nu}(g,f^i;\beta_n^{(i)})&=&0\label{eomcen}\,,\\
\mathcal{G}_{\mu\nu}(f^i)+\frac{m^2}{\alpha_i^2}V_{\mu\nu}(f^i, g;\beta^{(i)}_{4-n})&=&0\,,\qquad i=1,\hdots,N\,.\label{eomsat}
\eeqn
\end{subequations}
Here, $\mathcal{G}_{\mu\nu}(g)=R_{\mu\nu}(g)-\frac{1}{2}\gmn R(g)$ denotes the Einstein tensor and the contributions coming from the interaction potential are of the following form~\cite{Hassan:2011vm},
\beqn\label{poteom}
V_{\mu\nu}(g,f;\beta_n)&=& g_{\mu\rho}\sum_{n=0}^3(-1)^n\beta_n (Y_{(n)})^\rho_{~\nu}(S)\,.
\eeqn
They contain matrix valued functions $(Y_{(n)})^\rho_{~\nu}(S)$ defined as,
\beqn
(Y_{(n)})^\rho_{~\nu}(S)&=&\sum_{k=0}^n (-1)^k e_k(S) (S^{n-k})^\rho_{~\nu}\,,\qquad S^\rho_{~\nu}=\big(\sqrt{g^{-1}f}\,\big)^\rho_{~\nu}\,,
\eeqn
and also enter the bimetric equations~(\ref{eombt}).
Hence, the above multimetric equations are simply a straightforward generalisation of the well-known bimetric case as well. The structure of interactions is such that $\gmn$ appears in all $(N+1)$ sets of equations, while each $\fmn^i$ shows up only in its own equations and in those for $\gmn$.

\subsection{Action with chain couplings}
The other type of coupling for $N$ tensor fields which does not introduce the Boulware-Deser ghosts has the form,
\beqn\label{actchain}
S_\mathrm{chain}[g^i]=m_g^2\int\dd^4x\Bigg(\sum_{i=1}^N\alpha^2_i \sqrt{g^i}~R(g^i)
-2m^2\sum_{i=2}^{N}V(g^i,g^{i-1};\beta_n^{(i)})\Bigg)\,.
\eeqn
This action contains $N$ Planck masses $\alpha_i m_g$.
The interactions are again simple copies of~(\ref{bimpot}) with $(N-1)$ sets of parameters $\beta_n^i$. Each tensor field $\gmn^i$ with $i=2,\hdots, N-1$ interacts with two others, $\gmn^{i\pm1}$. The two fields at the end of the chain, $\gmn^1$ and $\gmn^N$ only possess interactions with one other field. Observe that, again, it is not possible to close the loop and let $\gmn^1$ interact with $\gmn^N$. Also all types of other additional interactions among the $\gmn^i$ are forbidden because they would create loops and render the theory inconsistent. Note also that the parameters $\beta_0^i$ are degenerate with $\beta_4^{i-1}$, but in order to keep the description more symmetric, we allow for this degeneracy here. The chain coupling is displayed in Figure~3. 
   \begin{figure}[h]
    \begin{center}
    \includegraphics[width=90pt]{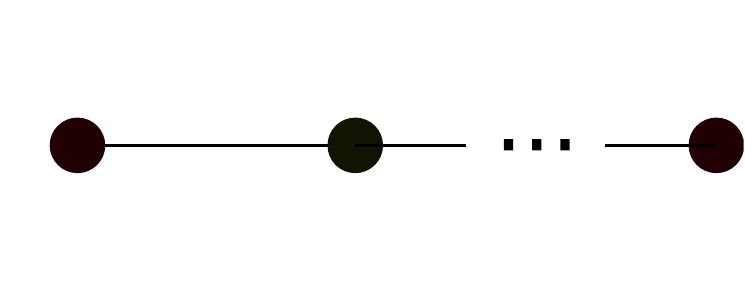}
    \caption{Graph for a chain coupling.}
    \end{center}\label{figcentchaint}
    \end{figure}
\\
The structure of the chain coupling does not possess a simple invariance under a set of individual interchange symmetries as in~(\ref{intsym}). It is however however symmetric under a reflection of the whole chain which interchanges $\gmn^1$ with $\gmn^N$, $\gmn^2$ with $\gmn^{N-1}$, etc. Explicitly, this invariance reads,
\beqn
\alpha_i^2\gmn^i\longleftrightarrow\alpha_{N-i}^2\gmn^{N-i}\,,\qquad
\frac{\beta_n^i}{\alpha^n_i}\longleftrightarrow\frac{\beta_{4-n}^{N-1}}{\alpha^{4-n}_{N-1}}\,, \qquad
\forall\, i=1,\hdots N\,.
\eeqn
This is only one interchange symmetry in total whereas, as we saw above, the structure of the centre coupling possesses $(N-1)!$ such invariances. In this sense, the centre coupling possesses a higher degree of symmetry. As a consequence, the analysis of multimetric properties, such as the derivation of maximally symmetric background solutions and the mass spectrum, tends to be more difficult for the chain coupling.

The equations of motion for the chain coupling read,
\begin{subequations}\label{eomchain}
\beqn
\mathcal{G}_{\mu\nu}(g^i)+\frac{m^2}{\alpha_i^2}\Big( V_{\mu\nu}(g^i,g^{i-1};\beta_n^{(i)})
+ V_{\mu\nu}(g^{i}, g^{i+1};\beta_{4-n}^{(i+1)})\Big)
&=&0\,,
\\
\text{for}~~i&=&2,3,\hdots N-1\,,\nn\\
\mathcal{G}_{\mu\nu}(g^1)+\frac{m^2}{\alpha_1^2}V_{\mu\nu}(g^1, g^2;\beta_{4-n}^{(2)})&=&0\,,\\
\mathcal{G}_{\mu\nu}(g^N)+\frac{m^2}{\alpha_N^2} V_{\mu\nu}(g^N,g^{N-1};\beta_{n}^{(N)})&=&0\,.
\eeqn
\end{subequations}
These are again expressed in terms of the Einstein tensors $\mathcal{G}_{\mu\nu}$ and the potential contributions are of the same functional form as in~(\ref{poteom}).
The first $(N-2)$ sets of equations arise from varying the action with respect to the fields in the middle of the chain, whereas the last two come from the metrics that sit on its ends. Note that each metric $\gmn^i$ only appears in the equations of its direct neighbours.

\subsection{A few additional remarks}

Before we get started to analyse multimetric interactions, let us note that the most general consistent multimetric actions combine the chain with the centre couplings, as depicted in Figure~4. Each of the legs may itself be coupled to additional metrics through chain or centre interactions.
   \begin{figure}[h]
    \begin{center}
    \includegraphics[width=120pt]{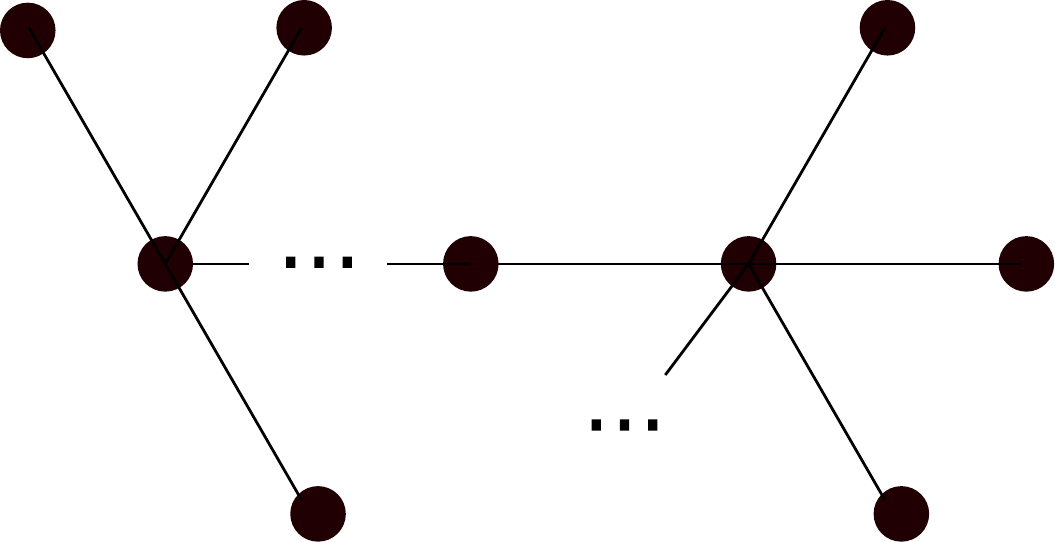}
    \caption{A consistent combination of centre and chain couplings.}
    \end{center}\label{fig4}
    \end{figure}
    \\
Observe also that in bimetric theory with two metrics the two types of couplings coincide. This is still true for the consistent action of three interacting metrics: One can imagine their couplings either as a chain of three fields or as two fields interacting with one in the centre. Since this ``trimetric" case is already nontrivial and captures the structure of both types of couplings, we will frequently use it as an example for the general case of an arbitrary number of interacting tensor fields. In fact, all our definite results will be obtained for the case of three interacting metrics.

In the majority of this paper, we will focus on the centre coupling because it is significantly simpler to analyse. Analogous computations for the chain couplings turn out to be much more complex and we will comment on them in appendix~\ref{app:chcoupl}. For the same reason, we will not discuss combined couplings like those in Figure~4 in this work. 
Moreover, we will concentrate on vacuum solutions to the multimetric equations of motion and therefore neglect couplings to matter. We restrict the analysis here to $d=4$ spacetime dimensions; note that in higher dimensions additional kinetic terms (which are topological or vanish in $d=4$) need to be considered~\cite{Paulos:2012xe, Hassan:2012rq}.

\section{Maximally symmetric solutions and mass spectrum}\label{sec:pbgms}

We will now derive the proportional background solutions in multimetric theories and discuss the mass spectrum of small perturbations around them. Our method is a direct generalisation of the bimetric analysis performed in Ref.~\cite{Hassan:2012wr, Hassan:2012rq}.

Before we get started let us make a brief comment on an interesting feature of the bimetric equations of motion~(\ref{eombt}), pointed out in Ref.~\cite{Hassan:2014vja}. Namely, as soon as one of the metrics, say $\gmn$, is assumed to be an Einstein metric with $R_{\mu\nu}(g)=\lambda_g\gmn$, then the equations imply that the second metric has to be Einstein, too, i.e.~$R_{\mu\nu}(f)=\lambda_f\fmn$. In the case where two out of the three interaction parameters $\beta_n$ vanish, the equations furthermore imply that then the two metrics are proportional to each other, $\fmn=c^2\gmn$, and that the cosmological constants satisfy $\lambda_g=c^2\lambda_f$. For general interaction parameters, the two background curvatures need not be related in this way and the metrics can be non-proportional Einstein backgrounds. It is quite straightforward to generalise these results to multimetric interactions. The outcome is that if all but one metric in the equations for the centre coupling~(\ref{eomcenter}) or for the chain coupling~(\ref{eomchain}) are assumed to be Einstein, then the last one has to be an Einstein metric as well.  In general, the corresponding solutions will be non-proportional Einstein backgrounds, which we shall not discuss any further in this work. Instead we focus on the case of proportional metrics which allow for a well-defined mass spectrum at the perturbative level.

\subsection{Three metrics}
In order to get a feeling for the general structure, which is identical for any number of metrics interacting in a centre coupling, we start by discussing the simplest nontrivial case with two tensor fields coupled to a third one in the centre.
Hence in this section we will work with the simplest generalisation of the ghost-free bimetric action,
\beqn\label{trimetricact}
S[g,f,h]=m_g^2\int\dd^4x\Bigg( \sqrt{g}~R(g)&+&\alpha_f^2 \sqrt{f}~R(f)+\alpha_h^2 \sqrt{h}~R(h)\nn\\
&-&2m^2V(g,f; \beta_n^f)-2m^2V(g,h; \beta_n^h)\Bigg)\,.
\eeqn
It includes three metrics $\gmn$, $\fmn$ and $\hmn$, whose interactions involve two sets of parameters $\beta_n^f$ and $\beta_n^h$ and read,
\beqn
V(g,f; \beta_n^f)=\sqrt{g}\sum_{n=0}^4\beta^{f}_n e_n\big(\sqrt{g^{-1}f}\,\big)\,,\qquad
V(g,h; \beta_n^h)=\sqrt{g}\sum_{n=0}^4\beta^{h}_n e_n\big(\sqrt{g^{-1}h}\,\big)\,.
\eeqn
In these, $\beta_0^f$ and $\beta_0^h$ both multiply the same term proportional to $\sqrt{g}$ and are therefore degenerate.
As remarked upon earlier, the trimetric potential can either be interpreted as a chain of three fields with $\fmn$ and $\hmn$ sitting at the end of the chain, or we can regard it as a centre coupling with $\gmn$ being the central metric. The trimetric theory thus corresponds to the action~(\ref{actcentre}) with $N=2$ and to the action~(\ref{actchain}) with $N=3$.

\subsubsection{Proportional backgrounds}\label{sec:pbgthree}

In terms of the respective Einstein tensors $\mathcal{G}_{\mu\nu}$, the equations of motion are,
\begin{subequations}\label{trieq}
\beqn
\mathcal{G}_{\mu\nu}(g)+m^2\Big(V_{\mu\nu}(g,f;\beta_n^f)+V_{\mu\nu}(g,h;\beta_n^h)\Big)&=&0\,,\\
\mathcal{G}_{\mu\nu}(f)+\frac{m^2}{\alpha_f^2}V_{\mu\nu}(f, g;\beta_{4-n}^f)=0\,,\qquad
\mathcal{G}_{\mu\nu}(h)+\frac{m^2}{\alpha_h^2}V_{\mu\nu}(h, g;\beta_{4-n}^h)&=&0\,,
\eeqn
\end{subequations}
where the $V_{\mu\nu}$ are given through the matrix functions $Y^{(n)}$ as in~(\ref{poteom}). In the simplest class of solutions to these equations the three metrics are all proportional to each other.
In order to analyse these backgrounds, we make the following ansatz,
\beqn
\fmn=c_f^2\gmn\,,\qquad
\hmn=c_h^2\gmn\,,
\eeqn
with proportionality constants $c^2_f$ and $c^2_h$ that will be determined by the equations. As in the bimetric case, the Bianchi constraints obtained from taking the divergences of the equations, would imply that $c^2_f$ and $c^2_h$ are constant even if we had assumed a coordinate dependence to start with.
On the proportional ansatz, the equations simply reduce to,
\begin{subequations}\label{syspbg}
\beqn
\mathcal{G}_{\mu\nu}(g)+\gmn\Big(\Lambda(\beta^f_n,c_f)+\Lambda(\beta^h_n,c_h)\Big)&=&0\,,\\
\mathcal{G}_{\mu\nu}(c_f^2g)+\gmn\tilde{\Lambda}(\beta^f_n,c_f, \alpha_f)&=&0\,,\\
\mathcal{G}_{\mu\nu}(c_h^2g)+\gmn\tilde{\Lambda}(\beta^h_n,c_h,\alpha_h)&=&0\,,
\eeqn
\end{subequations}
where the cosmological constants are functions of the parameters in the action as well as the proportionality constants in the ansatz,
\begin{subequations}\label{ccdef}
\beqn
\Lambda(\beta_n, c)&=&m^2\big(\beta_0+3\beta_1c+3\beta_2c^2+\beta_3c^3\big)\,, \\
\tilde{\Lambda}(\beta_n, c, \alpha)&=&\tfrac{m^2}{\alpha^2c^2}\Big(\beta_1c+3\beta_2c^2+3\beta_3c^3+\beta_4c^4\Big)\,.
\eeqn
\end{subequations}
The trimetric equations of motion thus become three sets of Einstein's equations and accordingly the solutions are maximally symmetric Einstein backgrounds. As already stated above, applying the same arguments as for bimetric theory in Ref.~\cite{Hassan:2014vja}, it is easy to see that, for general interaction parameters, the trimetric theory also allows for additional maximally symmetric solutions which are non-proportional. We concentrate on the proportional backgrounds because they are the only ones that allow for a definite mass spectrum in terms of spin-2 states. 

Since the Einstein tensor is scale invariant, $\mathcal{G}_{\mu\nu}(c^2g)=\mathcal{G}_{\mu\nu}(g)$ for any constant $c^2$, we can build differences of the above equations and derive two conditions on the cosmological constants,
\beqn\label{req}
\Lambda(\beta^f_n,c_f)+\Lambda(\beta^h_n,c_h)=\tilde{\Lambda}(\beta^f_n,c_f, \alpha_f)\qquad \text{and}\qquad \tilde{\Lambda}(\beta^f_n,c_f, \alpha_f)=\tilde{\Lambda}(\beta^h_n,c_h,\alpha_h)\,.
\eeqn
Consistent background solutions need to satisfy these polynomial equations for $c_f^2$ and $c_h^2$. Hence, the two conditions determine the values for the two proportionality constants in terms of the parameters in the trimetric action and this fully specifies the solution.

\subsubsection{Mass eigenstates}\label{sec:masstm}

Owing to the presence of the square-root matrix in the interaction potential, the linearised equations of motion around general backgrounds have a rather complicated form~\cite{Bernard:2015mkk, Cusin:2015tmf, Bernard:2015uic}. For the proportional solutions of the previous subsection, however, the structure simplifies and deriving the spectrum of mass eigenstates is straightforward.

We first perturb the background metrics according to $\gmn=\bar{g}_{\mu\nu}+\delta\gmn$,  $\fmn=c_f^{2}\bar{g}_{\mu\nu}+\delta\fmn$ and  $\hmn=c_h^{2}\bar{g}_{\mu\nu}+\delta\hmn$. 
Then we linearise the trimetric equations~(\ref{trieq}) in the perturbations which gives,
\begin{subequations}\label{systemlincentre}
\beqn
\bar{\mathcal{E}}^{~~\rho\sigma}_{\mu\nu}\delta g_{\rho\sigma} - m^2\mathcal{A}_f\big( \delta S^f_{\mu\nu}-\delta S^f\bar{g}_{\mu\nu}\big)
-m^2\mathcal{A}_h\big( \delta S^h_{\mu\nu}-\delta S^h\bar{g}_{\mu\nu}\big)
&=&0\,,\\
\bar{\mathcal{E}}^{~~\rho\sigma}_{\mu\nu}\delta  f_{\rho\sigma} + \tfrac{m^2}{\alpha_f^2}\mathcal{A}_f\big( \delta S^f_{\mu\nu}-\delta S^f\bar{g}_{\mu\nu}\big)&=&0\,,\\
\bar{\mathcal{E}}^{~~\rho\sigma}_{\mu\nu}\delta  h_{\rho\sigma} + \tfrac{m^2}{\alpha_h^2}\mathcal{A}_h\big( \delta S^h_{\mu\nu}-\delta S^h\bar{g}_{\mu\nu}\big)&=&0\,,\label{eomdh}
\eeqn
\end{subequations}
Here we have expressed the contributions coming from the interaction potential in terms of the following two linear combinations of fluctuations,
\beqn
\delta S^f_{\mu\nu}&\equiv&\frac{1}{2c_f}\big(\delta  f_{\mu\nu}-c_f^2\delta\gmn\big)\,,\qquad
\delta S^h_{\mu\nu}\equiv\frac{1}{2c_h}\big(\delta  h_{\mu\nu}-c_h^2\delta\gmn\big)\,.
\eeqn
They enter the equations in the Fierz-Pauli combination $(\delta S_{\mu\nu}-\delta S\bar{g}_{\mu\nu})$, as appropriate for a spin-2 mass term, c.f. equation~(\ref{linbimeq}).
The kinetic structure is captured by the following two-derivative operator, in which we have also included all cosmological constant contributions coming from kinetic and potential terms,
\beqn\label{kinop}
\bar{\mathcal{E}}^{~~\rho\sigma}_{\mu\nu}\delta g_{\rho\sigma}= \tfrac{1}{2}\Big(
	\delta^\rho_{~\mu}\delta^\sigma_{~\nu}\bar\nabla^2
	&-&\delta_{~\nu}^{\sigma}\bar\nabla_\mu\bar\nabla^\rho
	-\delta_{~\mu}^{\sigma}\bar\nabla_\nu\bar\nabla^\rho
	+\bar{g}_{\mu\nu} \bar\nabla^\sigma\bar\nabla^\rho\nn\\
	&+&\bar{g}^{\rho\sigma}\bar\nabla_\mu\bar\nabla_\nu
	-\bar{g}_{\mu\nu}\bar{g}^{\rho\sigma}\bar\nabla^2
	-2\Lambda\delta^\rho_{~\mu}\delta^\sigma_{~\nu}
	+\Lambda\bar{g}_{\mu\nu}\bar{g}^{\rho\sigma}\Big)\delta g_{\rho\sigma}
	\,.
\eeqn
This operator is given in terms of $\bar{g}_{\mu\nu}$ in all three sets of equations. In addition to their appearance in the background cosmological constant~(\ref{ccdef}), the interaction parameters enter the linearised equations in the following combinations,
\beqn
\mathcal{A}_f&\equiv&  \beta^{f}_1+2c_f\beta^{f}_2+c_f^2\beta^{f}_3\,,\qquad
\mathcal{A}_h\equiv  \beta^{h}_1+2c_h\beta^{h}_2+c_h^2\beta^{h}_3\,.
\eeqn
Recall that in all expressions the proportionality constants $c_f$ and $c_h$ should be regarded as functions of trimetric parameters, determined by the background conditions in~(\ref{req}).
The system~(\ref{systemlincentre}) is a direct generalisation of the bimetric case, which formally corresponds to setting $\mathcal{A}_h$ to zero and neglecting the equation of motion~(\ref{eomdh}) for $\delta\hmn$.
 
In order to find the mass eigenstates, we need to build linear combinations of the fluctuations in such a way that each of them satisfies a decoupled set of Fierz-Pauli equations as in~(\ref{linbimeq}). By inspecting the system (\ref{systemlincentre}), we see that this task is equivalent to diagonalising the following $3\times 3$ mass matrix,
\beqn
\mathcal{M}=
m^2\begin{pmatrix}
\mathcal{A}_fc_f +\mathcal{A}_hc_h ~&~-\frac{\mathcal{A}_fc_f}{\alpha_f^2} ~&~-\frac{\mathcal{A}_hc_h}{\alpha_h^2}\\
-\frac{\mathcal{A}_f}{c_f}~&~\frac{\mathcal{A}_f}{\alpha_f^2c_f} ~&~ 0\\
 -\frac{\mathcal{A}_h}{c_h}~&~ 0 ~&~ \frac{\mathcal{A}_h}{\alpha_h^2c_h}
\end{pmatrix}\,,
\eeqn 
whose eigenvalues will correspond to the masses of the spin-2 states.
Deriving the characteristic equation for the eigenvalues $\mu^2$ of $\mathcal{M}$ is straightforward,
\beqn\label{chareq}
0=\det \big(\mu^2\mathbb{1}-\mathcal{M}\big)=\mu^2\Big(\mu^4+p\mu^2+q\Big)\,.
\eeqn
Here we have used the abbreviations,
\beqn
p&=&-\mathcal{A}_f\Big(c_f+\frac{1}{\alpha_f^2c_f}\Big)m^2 -\mathcal{A}_h\Big(c_h+\frac{1}{\alpha_h^2c_h}\Big)m^2
\,,\\
q&=& \frac{\mathcal{A}_f\mathcal{A}_h}{\alpha_f^2\alpha_h^2c_fc_h}\Big(1+\alpha_h^2c_h^2+\alpha_f^2c_f^2\Big)m^4\,.
\eeqn
The mass eigenvalues correspond to the roots of (\ref{chareq}), which are directly obtained as,
\beqn\label{FPmasses}
\mu_0^2=0\,,\qquad
\mu_\pm^2=-\tfrac{p}{2}\pm\sqrt{\left(\tfrac{p}{2}\right)^2-q}\,.
\eeqn
As expected, the spectrum contains one massless spin-2 mode and two massive ones. The masses of the latter are complicated functions of the trimetric parameters and generically different from each other.

The eigenvectors $\vec{v}(\mu^2)$ corresponding to the above mass values are obtained from solving the eigenvalue equation, $\mathcal{M}\cdot \vec{v}=\mu^2\vec{v}$. This leads to the following result for the components of $\vec{v}$, up to an overall normalisation,
\beqn\label{masseigenvtm}
 v_1=1 \,,\qquad
 v_2=\frac{\mathcal{A}_f\alpha_f^2}{\mathcal{A}_f-\alpha_f^2c_f\mu^2/m^2}\,,\qquad
 v_3=\frac{\mathcal{A}_h\alpha_h^2}{\mathcal{A}_h-\alpha_h^2c_h\mu^2/m^2}\,,
\eeqn
where we have assumed $\mathcal{A}_f\neq\alpha_f^2c_f\mu^2/m^2$ and $\mathcal{A}_h\neq\alpha_h^2c_h\mu^2/m^2$, which is true for generic parameters in the action. The exception will be discussed below.
The massless spin-2 eigenstate $\delta G_{\mu\nu}$ and the two massive states $\delta M^\pm_{\mu\nu}$ thus correspond to the following linear combinations of the original variables,
\beqn
\delta G_{\mu\nu}&=&\delta\gmn+\alpha_f^2\delta  f_{\mu\nu}+\alpha_h^2\delta  h_{\mu\nu}\,,\\
\delta M^+_{\mu\nu}&=&\delta\gmn+\frac{\mathcal{A}_f\alpha_f^2}{\mathcal{A}_f-\alpha_f^2c_f\mu_+^2/m^2}\delta  f_{\mu\nu}+\frac{\mathcal{A}_h\alpha_h^2}{\mathcal{A}_h-\alpha_h^2c_h\mu_+^2/m^2}\delta  h_{\mu\nu}\,,\\
\delta M^-_{\mu\nu}&=&\delta\gmn+\frac{\mathcal{A}_f\alpha_f^2}{\mathcal{A}_f-\alpha_f^2c_f\mu_-^2/m^2}\delta  f_{\mu\nu}+\frac{\mathcal{A}_h\alpha_h^2}{\mathcal{A}_h-\alpha_h^2c_h\mu_-^2/m^2}\delta  h_{\mu\nu}\,.
\eeqn
In summary, we have found that the mass spectrum of three interacting metrics can be explicitly computed and the linearised equation are diagonalised into one equation for a massless and two equations for massive spin-2 fields,
\beqn
\bar{\mathcal{E}}^{~~\rho\sigma}_{\mu\nu}\delta G_{\rho\sigma}&=&0\,,\\
\bar{\mathcal{E}}^{~~\rho\sigma}_{\mu\nu}\delta  M^\pm_{\rho\sigma} - \frac{\mu_\pm^2}{2}\big( \delta M^\pm_{\mu\nu}-\delta M^\pm\bar{g}_{\mu\nu}\big)&=&0\,,
\eeqn
with Fierz-Pauli masses $\mu^2_\pm$ given in~(\ref{FPmasses}).

\paragraph{A simplification:} 
In order to derive the above mass eigenstates, we had to make a restrictive assumption on the eigenvalues which shall now be relaxed.
That is to say, we now assume the existence of an eigenvalue $\mu^2$ such that either $\mathcal{A}_f=\alpha_f^2c_f\mu^2/m^2$ or $\mathcal{A}_h=\alpha_h^2c_h\mu^2/m^2$, in which case the generic solutions in (\ref{masseigenvtm}) do not exist. 
It is then straightforward to show that for either of these values of $\mu^2$ the characteristic equation~(\ref{chareq}) implies that the parameters must satisfy $\frac{1}{\alpha_f^2c_f}\mathcal{A}_f=\frac{1}{\alpha_h^2c_h}\mathcal{A}_h\equiv\mathcal{A}$. As a consequence, the mass eigenstates assume the following simple forms,
\beqn\label{specialstates}
\delta G_{\mu\nu}&=&\delta\gmn+\alpha_f^2\delta  f_{\mu\nu}+\alpha_h^2\delta  h_{\mu\nu}\,,\\
\delta M^1_{\mu\nu}&=& \delta  f_{\mu\nu}-\frac{c_f^2}{c_h^2}\delta  h_{\mu\nu}\,,\qquad
\delta M^2_{\mu\nu}= \alpha_f^2\delta  f_{\mu\nu}+\alpha_h^2\delta  h_{\mu\nu}-(\alpha_f^2c_f^2+\alpha_h^2c_h^2)\delta\gmn\,.
\eeqn
Their corresponding mass eigenvalues are,
\beqn
\mu^2_0=0\,,
\qquad
\mu^2_1=\mathcal{A}m^2
\,,\qquad
\mu_2^2= \mathcal{A}m^2\big(1+\alpha_f^2c_f^2+\alpha_h^2c_h^2\big)\,.
\eeqn
Clearly, this very simple solution only exists for a specific choice of interaction parameters tuned to ensure $\frac{1}{\alpha_f^2c_f}\mathcal{A}_f=\frac{1}{\alpha_h^2c_h}\mathcal{A}_h$. An example for such a choice would be to set $\beta_f=\frac{\alpha_f^2}{\alpha_h^2}\beta_h$ and focus on the case $c_f=c_h$, which is now a solution to the second condition in~(\ref{req}). 
We will encounter this very symmetric situation again in section~\ref{sec:carbthree}.

\subsection{Centre coupling}
In the following we generalise the analysis of the trimetric theory to $N$ fields $\fmn^i$ coupled to a central metric $\gmn$. The corresponding action was given in~(\ref{actcentre}). The results for the background solutions and mass eigenstates turn out to be straightforward generalisations of those found for three metrics. The mass eigenvalues are roots of an $N$th order polynomial equation and can easily be computed by hand in a simplified example.

\subsubsection{Proportional backgrounds}\label{sec:propbgcentre}

In the equations of motion~(\ref{eomcenter}) for $(N+1)$ metrics in a centre coupling, we make an ansatz for proportional backgrounds which reads,
\beqn
f^i_{\mu\nu}=c_i^2\gmn\,,
\qquad i=1,\hdots ,N\,,
\eeqn
with proportionality constants $c_i^2$. On this ansatz the equations reduce to $(N+1)$ copies of Einstein's equations,
\beqn\label{bgeqcc}
\mathcal{G}_{\mu\nu}(g)+\gmn\sum_{k=1}^N \Lambda(\beta_n^{(k)}, c_k)=0\,,\qquad
\mathcal{G}_{\mu\nu}(f^i)+\gmn\,\tilde\Lambda(\beta_n^{(i)}, c_i, \alpha_i)=0\,.
\eeqn
The cosmological constants are of the same functional form as in (\ref{ccdef}). Using again the scale invariance of the Einstein tensor, $\mathcal{G}_{\mu\nu}(c_i^2\gmn)=\mathcal{G}_{\mu\nu}(\gmn)$, we then obtain the following consistency conditions on the proportionality constants,
\beqn
\sum_{i=1}^N \Lambda(\beta_n^{(i)}, c_i)=\tilde\Lambda(\beta_n^{(j)}, c_j, \alpha_j)\,,\qquad  j=1,\hdots,N\,.
\eeqn
These constitute $N$ polynomial equations which determine all $c_i^2$ and thus fully specify the solution. From now on we can think of $c_i$ as given in terms of multimetric parameters.

\subsubsection{Mass eigenstates}\label{sec:masscentre}

As in the trimetric case, we proceed by studying the linear theory around proportional backgrounds. Decomposing $\gmn=\bar{g}_{\mu\nu}+\delta \gmn$ and $f^i_{\mu\nu}=c_i^2\bar{g}_{\mu\nu} +\delta \fmn^i$, we obtain the linearised equations for the metric perturbations as,
\beqn
\bar{\mathcal{E}}^{~~\rho\sigma}_{\mu\nu}\delta g_{\rho\sigma} - m^2\sum_{i=1}^N\mathcal{A}_i\big( \delta S^i_{\mu\nu}-\delta S^i\bar{g}_{\mu\nu}\big)&=&0\,,\\
\bar{\mathcal{E}}^{~~\rho\sigma}_{\mu\nu}\delta  f^i_{\rho\sigma} + \frac{m^2}{\alpha_i^2}\mathcal{A}_i\big( \delta S^i_{\mu\nu}-\delta S^i\bar{g}_{\mu\nu}\big)&=&0\,,\qquad i=1,\hdots, N\,.
\eeqn
Here, for $i=1,\hdots, N$, we have defined the linear combinations of fluctuations,
\beqn
\delta S^i_{\mu\nu}\equiv\frac{1}{2c_i}\big(\delta  f^i_{\mu\nu}-c_i^2\delta\gmn\big)\,,
\eeqn
which appear in the contributions from the interaction potential with Fierz-Pauli structure. The kinetic operator is the same as in (\ref{kinop}), defined in terms of the background metric $\bar{g}_{\mu\nu}$ and the cosmological constant of the background equations~(\ref{bgeqcc}). The interaction parameters enter the linearised equations through the cosmological constant as well as the familiar combinations,
\beqn
\mathcal{A}_i\equiv  \beta^{(i)}_1+2c_i\beta^{(i)}_2+c_i^2\beta^{(i)}_3\,, \qquad i=1,\hdots, N\,.
\eeqn
In order to find the spin-2 mass eigenstates we now need to solve the eigenvalue equation for an $(N+1)\times(N+1)$ matrix with the following structure,
\beqn\label{massmatrix}
\mathcal{M}=
m^2\begin{pmatrix}
\sum_{i=1}^N\mathcal{A}_ic_i  ~&~-\frac{\mathcal{A}_1c_1}{\alpha_1^2} ~&~-\frac{\mathcal{A}_2c_2}{\alpha_2^2} &~\hdots ~&~-\frac{\mathcal{A}_Nc_N}{\alpha_N^2}\\
-\frac{\mathcal{A}_1}{c_1}~&~\frac{\mathcal{A}_1}{\alpha_1^2c_1} ~&~0~&~\hdots~& 0\\
-\frac{\mathcal{A}_2}{c_2}~&~0~&~\frac{\mathcal{A}_2}{\alpha_2^2c_2} ~&~\hdots~&\vdots\\
\vdots~&\vdots&&\ddots&0\\
 -\frac{\mathcal{A}_N}{c_N}~&~ 0 ~&~\hdots~&~0~& \frac{\mathcal{A}_N}{\alpha_h^2c_N}
\end{pmatrix}\,.
\eeqn 
This is a straightforward computation, completely analogous to the trimetric case, and the mass eigenstates corresponding to eigenvalues $\mu_k^2$ turn out to be the following linear combinations of the original fluctuations,
\begin{subequations}
\beqn\label{masslessescentre}
\delta G_{\mu\nu}&=&\delta\gmn+\sum_{i=1}^N\alpha_i^2\delta  f^i_{\mu\nu}\,,\label{masslessescentre}\\
\delta M^k_{\mu\nu}&=&\delta\gmn+\sum_{i=1}^N\frac{\mathcal{A}_i\alpha_i^2}{\mathcal{A}_i-\alpha_i^2c_i\mu_k^2/m^2}\delta  f^i_{\mu\nu}\,,
\qquad k=1,\hdots,N\,,\label{massescentre}
\eeqn
\end{subequations}
where the first combination $\delta G_{\mu\nu}$ has the eigenvalue $\mu_0^2=0$.
Note that in order to arrive at the expressions for the massive states, we had to assume $\mathcal{A}_i\neq\alpha_i^2c_i\mu_k^2/m^2$ for all~$i$ and all~$\mu^2_k$.
The above result implies that the equations can be diagonalised into one equation for a massless and $N$ equations for massive spin-2 fields,
\begin{subequations}\label{linmmeqcc}
\beqn
\bar{\mathcal{E}}^{~~\rho\sigma}_{\mu\nu}\delta G_{\rho\sigma}&=&0\,,\\
\bar{\mathcal{E}}^{~~\rho\sigma}_{\mu\nu}\delta  M^i_{\rho\sigma} - \frac{\mu_i^2}{2}\big( \delta M^i_{\mu\nu}-\delta M^i\bar{g}_{\mu\nu}\big)&=&0\,,\qquad i=1,\hdots, N\,.
\eeqn
\end{subequations}
The Fierz-Pauli masses $\mu_k^2$ are the solutions to the characteristic equation, $\det(\mu^2\mathbb{1}-\mathcal{M})=0$, of the $(N+1)\times(N+1)$ matrix $\mathcal{M}$. Since it is straightforward to isolate the eigenvalue $\mu_0^2=0$ and divide the characteristic equation by $\mu^2$, the remaining eigenvalues correspond to the roots of an $N$th order polynomial.

\paragraph{A simplification:} In the case with $(N+1)$ metrics it is not obvious that assuming one eigenvalue to equal one of the $\frac{1}{\alpha_i^2c_i}\mathcal{A}_i$ will automatically imply that $\frac{1}{\alpha_i^2c_i}\mathcal{A}_i=\frac{1}{\alpha_j^2c_j}\mathcal{A}_j$ for all $i,j$. In the trimetric case this followed from the characteristic equation, which now has a more complicated form and provides just one constraint that most likely cannot fix all $\frac{1}{\alpha_i^2c_i}\mathcal{A}_i$ to be equal. 
Nevertheless, let us assume here that $\frac{1}{\alpha_i^2c_i}\mathcal{A}_i=\frac{1}{\alpha_j^2c_j}\mathcal{A}_j\equiv\mathcal{A}$ for all $i,j$ and study this restrictive parameter choice as an example for the cases excluded in the derivation of the mass eigenstates~(\ref{massescentre}).
The eigenstates are now simply the massless one, $\delta G_{\mu\nu}$ as in (\ref{masslessescentre}), along with $N$ massive fields given by the following simple expressions,
\beqn\label{masseigsimcen}
\delta M^i_{\mu\nu}&=& \delta  f^i_{\mu\nu}-\frac{c_{i}^2}{c_{i+1}^2}\delta  f^{i+1}_{\mu\nu}\,,\qquad i=1,\hdots N-1\,,\\
\delta M^N_{\mu\nu}&=& \sum_{i=1}^N\Big(\alpha_i^2\delta  f^i_{\mu\nu}-\alpha_i^2c_i^2\delta\gmn\Big)\,.
\eeqn
The corresponding mass eigenvalues take on the form,
\beqn
\mu_i^2&=&\mathcal{A}m^2
\,,\qquad i=1,\hdots N-1\,,\\
\mu_N^2&=& \mathcal{A}m^2\Big(1+\sum_{i=1}^{N}\alpha_i^2c_i^2\Big)\,.
\eeqn
We will encounter an application for this simplified example in section~\ref{sec:carbcentre}.

\section{Models with additional structure}\label{sec:mwas}

In this section we will derive particularly interesting parameter choices which allow the multimetric models to possess additional structure, possibly even an enhanced symmetry. Several different methods have been developed to identify such special models in massive gravity and bimetric theory~\cite{deRham:2012kf, Hassan:2012gz, Hassan:2012rq, Deser:2013uy, deRham:2013wv, Hassan:2013pca, Hassan:2015tba} and we will employ two of them here to study the case of $(N+1)$ tensor fields in a centre coupling. For three metrics, we demonstrate that both methods lead to the same set of models, which reduces to one unique model if interchange symmetry among the satellite metrics is imposed.

\subsection{Scale invariance of proportional backgrounds}\label{sec:scinv}

The first method that we shall apply has been developed in Ref.~\cite{Hassan:2012gz, Hassan:2012rq}. It is based on the proportional background solutions in bimetric theory and aims to single out models for which these backgrounds are invariant under constant scalings. The main motivation of those references was the linear gauge symmetry of a partially massless spin-2 field in de Sitter backgrounds and the question whether this symmetry could be extended to the nonlinear level. A necessary requirement for such an extension is the possibility of transferring the constant part of the linear gauge transformation from the fluctuations to the background, resulting in the mentioned scale invariance (see~\cite{Hassan:2012gz} for a detailed derivation of this criterion).
It should be emphasised that this requirement is necessary but not sufficient for the existence of a symmetry beyond the constant curvature backgrounds.

Since then, many other works have argued against the possible realisation of nonlinear partial masslessness~\cite{Deser:2013uy, Deser:2013gpa, deRham:2013wv, Joung:2014aba, Garcia-Saenz:2014cwa, Apolo:2016vkn}. It is therefore very likely that a theory involving only spin-2 degrees of freedom will not be able to deliver a description of interacting partially massless fields. On the other hand, the inclusion of additional fields could potentially circumvent the problems. A recent suggestion is based on the bimetric potential identified in~\cite{Hassan:2012gz} and includes an additional vector field~\cite{Apolo:2016ort}. Another interesting approach is to consider a full tower of higher-spin states~\cite{Brust:2016zns}. 
 
Here, we choose not to use partial masslessness as our main motivation but simply employ the known methods to look for additional structure. At this stage it is not obvious whether the multimetric potentials with these special properties will be helpful for further studying nonlinear interactions of partially massless fields. 
Nevertheless, as we shall see, the models that we identify in the following do propagate partially massless fields around their maximally symmetric solutions. The linear theory is therefore invariant under the additional gauge theory but we do not make any claims about possible nonlinear extensions in what follows.

\subsubsection{Three metrics}\label{sec:carbthree}

As before, we start with the illustrative example of three metrics for which the proportional background solutions with $\fmn=c_f^2\gmn$ and $\hmn=c_h^2\hmn$ were derived in section~\ref{sec:pbgthree}. Let us remind the reader that the equations of motion implied two conditions~(\ref{req}) on the cosmological constants whose expressions were given in~(\ref{ccdef}). These constraints read,
\beqn\label{req2}
\Lambda(\beta^f_n,c_f)+\Lambda(\beta^h_n,c_h)=\tilde{\Lambda}(\beta^f_n,c_f, \alpha_f)\qquad \text{and}\qquad \tilde{\Lambda}(\beta^f_n,c_f, \alpha_f)=\tilde{\Lambda}(\beta^h_n,c_h,\alpha_h)\,.
\eeqn
Usually, they determine the two proportionality constants $(c_f,c_h)$ and thereby fully specify the solution to the equations of motion.
We would now like to require the backgrounds to be invariant under a constant scale transformation. This means that one of $(c_f,c_h)$ must be undetermined by the equations such that it can be shifted in an arbitrary way and still result in a valid background configuration.\footnote{It is straightforward to see that two such invariances cannot exist since one of $(c_f,c_h)$ will always be determined by the equations.} In other words, suppose that we have solved one of the above constraints for $c_h$ as a function of $c_f$ and that then the second constraint is satisfied identically and does therefore not determine $c_f$. In this case, any value for $c_f$ is a solution to the equations of motion and its shifts constitute an invariance of the proportional backgrounds.

In order to see which parameter configurations lead to scale invariance, we thus start by assuming a general functional dependence $c_h=F(c_f)$ and demand that $c_f$ be undetermined. On this ansatz, the explicit form of the second constraint in~(\ref{req2}) is,
\beqn
\frac{\alpha_h^2 F(c_f)}{\alpha_f^2c_f}\Big(\beta^f_1+3\beta_2^fc_f+3\beta^{f}_3c_f^2+\beta_4^fc_f^3\Big)
=\beta^h_1+3\beta_2^hF(c_f)+3\beta^{h}_3F(c_f)^2+\beta_4^hF(c_f)^3\,.
\eeqn
For general parameters, we would have to solve this cubic polynomial equation for $F(c_f)$ which is possible but not very illuminating. The crucial point is that the final equation for $c_f$ (corresponding to the first constraint in~(\ref{req2})) will generically not be of polynomial form and therefore contain and determine $c_f$, regardless of the parameter choice. However, the non-polynomial contributions are removed if and only if we choose,
\beqn\label{beta13}
\beta_1^f=\beta_1^h=\beta_3^f=\beta_3^h=0\,.
\eeqn 
In this case, the above constraint straightforwardly yields the simple quadratic solution,
\beqn
 F(c_f)^2=\frac{\alpha_h^2}{\beta_4^h\alpha_f^2}\Big(3\beta_2^f+\beta_4^fc_f^2\Big)-\frac{3\beta_2^h}{\beta_4^h}
\,.
\eeqn
Next, we use this expression to replace $c_h^2$ in the second constraint in~(\ref{req2}). This gives,
\beqn
\alpha_f^2\beta_4^h\beta_0^f+9\alpha_h^2\beta_2^h\beta_2^f-9\alpha_f^2(\beta_2^h)^2-3\beta_2^f\beta_4^h
+\Big(3\alpha_f^2\beta_2^f\beta_4^h+3\alpha_h^2\beta_2^h\beta_4^f-\beta_4^f\beta_4^h\Big)c_f^2=0\,,
\eeqn
where we have already set $\beta_0^h=0$ to remove the redundancy with $\beta_0^f$.
In general, this condition contains and therefore determines $c_f$. The proportionality constant remains undetermined only if it drops out of the equation. This finally requires the following quadratic combinations of parameters to vanish,
\begin{subequations}\label{betaprop}
\beqn
\alpha_f^2\beta_4^h\beta_0^f+9\alpha_h^2\beta_2^h\beta_2^f-9\alpha_f^2(\beta_2^h)^2-3\beta_2^f\beta_4^h&=&0\,,\\
3\alpha_f^2\beta_2^f\beta_4^h+3\alpha_h^2\beta_2^h\beta_4^f-\beta_4^f\beta_4^h&=&0\,.\label{seccontmec}
\eeqn
\end{subequations}
These conditions determine two out of the five remaining parameters $(\beta_0^f, \beta_2^f, \beta_4^f, \beta_2^h, \beta_4^h)$. Hence the resulting action is not unique but still possesses three free parameters (with one of them setting the overall interaction scale).

\paragraph {A simple example with additional symmetry:}
Suppose we had looked for a solution with $c_f=c_h\equiv c$, which reflects the symmetry of the centre interactions with respect to the satellite metrics. In this case, requiring $c$ to be undetermined would have resulted in,
\beqn\label{PMbetas2}
\beta^h_n=\Big(\frac{\alpha_h}{\alpha_f}\Big)^2\beta_n^f\,,
\qquad n=1,2,3,4\,,\nn\\
\alpha_f^2\beta^{f}_0=3\beta_2^f\,,\qquad 
3\big(\alpha_h^2+\alpha_f^2\big)\beta^{f}_2=\beta_4^f\qquad 
\beta^{f}_1=\beta^{f}_3=0\,,
\eeqn
which is a particular solution to the general constraints~(\ref{betaprop}).
Note that all but one interaction parameters $(\beta_n^f, \beta_n^h)$ are now fixed and the one remaining free parameter merely sets the overall scale of interactions. In other words, it comes as an overall factor in the trimetric potential and could be absorbed into the scale~$m^2$. 
The corresponding action is therefore uniquely determined with no more free parameters in its interaction potential.
Defining $\lambda\equiv 3\big(\alpha_h^2+\alpha_f^2\big)\beta_2^f m^2$, we can write it as,
\beqn
S[g,f,h]&=&m_g^2\int\dd^4x\Bigg( \sqrt{g}~R(g)
+\alpha_f^2 \sqrt{f}~\Big[R(f)-2\lambda\Big]
+\alpha_h^2 \sqrt{h}~\Big[R(h)-2\big(\tfrac{\alpha_h}{\alpha_f}\big)^2\lambda\Big]\nn\\
&~&\hspace{60pt}
-~2\beta_2^f m^2\sqrt{g}\Big[ \tfrac{3}{\alpha^2_f} +e_2\big(\sqrt{g^{-1}f}\,\big)+\big(\tfrac{\alpha_h}{\alpha_f}\big)^2e_2\big(\sqrt{g^{-1}h}\,\big) \Big]
\Bigg)\,.
\eeqn
It is now easy to verify that the constraints on the interaction parameters in~(\ref{PMbetas2}) imply an invariance of the action under the interchange,
\beqn
\alpha_h^2\hmn\longleftrightarrow\alpha_f^2\fmn\,.
\eeqn
Unlike the structural symmetry~(\ref{intsym}) which also requires a change in the parameters, this discrete transformation is a true invariance of the above action.

\paragraph{Mass spectrum:}

It is straightforward to verify that one of the massive spin-2 modes becomes partially massless for parameters that satisfy the relations (\ref{beta13}) and (\ref{betaprop}). The quickest way is to note that if we set,
\beqn
\mu^2=\frac{2}{3}\tilde{\Lambda}=\frac{1}{\alpha_f^2c_f}(3\beta_2^fc_f+\beta_4^fc_f^3)\,,
\eeqn
then the characteristic equation~(\ref{chareq}) for the eigenvalues $\mu^2$ of the mass matrix contains a factor of $(3\alpha_f^2\beta_2^f\beta_4^h+3\alpha_h^2\beta_2^h\beta_4^f-\beta_4^f\beta_4^h)$, which vanishes due to~(\ref{seccontmec}). This means that $\mu^2=\frac{2}{3}\tilde{\Lambda}$ is a mass eigenvalue for the above parameter choice and the corresponding linear equation possesses the partially massless gauge symmetry.

For the simplified case with parameter values specified by (\ref{PMbetas2}) we can easily demonstrate the presence of a partially massless mode explicitly. In this model, the parameter functions $\mathcal{A}_f$ and $\mathcal{A}_h$, which appear in the Fierz-Pauli masses~(\ref{FPmasses}), satisfy the following identity,
\beqn
\frac{1}{\alpha_h^2c_h}\mathcal{A}_h= \frac{1}{\alpha_f^2c_f}\mathcal{A}_f=\frac{2\beta^{f}_2}{\alpha_f^2}\,.
\eeqn
We observe that this constitutes an example for the exceptional case which yields a very simple mass spectrum, as we discussed at the end of section~\ref{sec:masstm}. 
The linearised equations around the proportional backgrounds thus read,
\beqn
\bar{\mathcal{E}}^{~~\rho\sigma}_{\mu\nu}\delta G_{\rho\sigma}&=&0\,,\\
\bar{\mathcal{E}}^{~~\rho\sigma}_{\mu\nu}\delta  M^i_{\rho\sigma} - \frac{\mu_i^2}{2}\big( \delta M^i_{\mu\nu}-\delta M^i\bar{g}_{\mu\nu}\big)&=&0\,, \qquad i=1,2\,.
\eeqn
where the mass eigenstates are the same as in~(\ref{specialstates}) with $c_f=c_h=c$.
Their corresponding mass eigenvalues are,
\beqn\label{masseshb}
\mu^2_0=0\,,
\qquad
\mu^2_1=\frac{2\beta^{f}_2}{\alpha_f^2}m^2
\,,\qquad
\mu_2^2= \frac{2\beta^{f}_2}{\alpha_f^2}m^2\big(1+\alpha_f^2c^2+\alpha_h^2c^2\big)\,.
\eeqn
Moreover, for our special parameter choice, the cosmological constant of the background, $R(\bar{g})_{\mu\nu}=\tilde{\Lambda}\bar{g}_{\mu\nu}$, assumes the following value,
\beqn\label{PMcctm}
\tilde{\Lambda}(\beta_n^f,c,\alpha_f)=\tfrac{3\beta^f_2m^2}{\alpha_f^2}\Big(1+\alpha_h^2c^2+\alpha_f^2c^2\Big)\,.
\eeqn
Comparing the mass eigenvalue $\mu_2^2$ to this cosmological constant, we find that they satisfy the relation,
\beqn
\mu_2^2=\frac{2}{3}\tilde{\Lambda}\,.
\eeqn
We have thus explicitly verified that the mass of the field $\delta M^2_{\mu\nu}$ is on the Higuchi bound and the linear theory exhibits the linear gauge symmetry of a partially massless spin-2 field.

Note that from (\ref{masseshb}) it follows that $\mu_1^2<\mu_2^2$. This implies that the second massive spin-2 field violates the Higuchi bound and is non-unitary~\cite{Higuchi:1986py}.\footnote{We thank an anonymous referee for bringing this to our attention.}
More generally, the results below will show that for $N$ fields the spectrum contains $(N-1)$ Higuchi ghosts. This problem therefore seems to be a direct consequence of demanding constant scale invariance and can only be avoided in pure bimetric theory with $N=1$.

\subsubsection{Centre coupling}\label{sec:carbcentre}

We now repeat the analysis of the previous subsection for the case of $N$ metrics $\fmn^i$ coupled to $\gmn$ in the centre. The proportional background solutions of this theory were derived in section~\ref{sec:propbgcentre} and we recall the background conditions,
\beqn\label{bgcondcc2}
\sum_{i=1}^N \Lambda(\beta_n^{(i)}, c_i)=\tilde\Lambda(\beta_n^{(j)}, c_j, \alpha_j)\,,\qquad \forall j=1,\hdots,N\,.
\eeqn
In general these determine all proportionality constants and here we look for special parameter choices for which one constant remains arbitrary.
In principle one can apply the procedure for three metrics iteratively and obtain all models with one undetermined proportionality constant. This will result in actions with $\beta_1^i=\beta_3^i=0$ for all $i$ and the other parameters will be constrained (but not fully determined) by a system of polynomial equations of order~$N$. We will not carry out this general computation explicitly.
Instead we shall focus on the class of simplified solutions for which all proportionality constant are equal, $c_i\equiv c~\forall i$, and require that $c$ is undetermined. This does not give all models with scale invariant backgrounds but, since it treats all satellite metrics $\fmn^i$ on equal footing, it is expected to result in the action with the highest amount of symmetry. 

For $c_i\equiv c$, the condition that the right-hand side of~(\ref{bgcondcc2}) is the same for all $j$ immediately forces us to fix,
\beqn\label{PMbetas1}
\beta^i_n=\left(\frac{\alpha_i}{\alpha_j}\right)^2\beta^j_n\qquad \text{for}~~n=1,2,3,4~~\text{and}~~\forall i,j\,,
\eeqn
in order to leave $c$ undetermined. Then, evaluating~(\ref{bgcondcc2}) for $j=1$ and requiring the coefficients of any power of $c$ to vanish yields,
\beqn\label{centcoupbeta}
\alpha_1^2\beta^{(1)}_0=3\beta_2^{(1)}\,,\qquad 
3\beta^{(1)}_2\sum_{k=1}^{N}\alpha_k^2=\beta_4^{(1)}\qquad 
\beta^{(1)}_1=\beta^{(1)}_3=0\,.
\eeqn
As in the trimetric example, these constraints fix all the $\beta_n^i$ in terms of one remaining parameter which parameterises the overall interaction scale. It is also easy to verify that the resulting action is invariant under the interchange of any two $\alpha_i^2\fmn^i$.
The above parameter values are therefore a direct generalisation of our result for three metrics.

\paragraph{Mass spectrum:}
Once more, the above parameter values correspond to the exceptionally simple mass spectrum which we discussed at the end of section~\ref{sec:masscentre}.  
In the corresponding linearised equations~(\ref{linmmeqcc}) the mass eigenstates are those of~(\ref{masseigsimcen}) with $c_i=c$ for all $i$. The masses of the massive fields are,
\beqn
\mu_i^2=\frac{2\beta^{f}_2}{\alpha_f^2}m^2
\,,\quad i=1,\hdots N-1\,,\qquad
\mu_N^2= \frac{2\beta^{f}_2}{\alpha_f^2}m^2\Big(1+c^2\sum_{k=1}^{N-1}\alpha_k^2\Big)
\eeqn
Comparing the last of these to the cosmological constant which reads,
\beqn\label{PMcccenter}
\tilde{\Lambda}=\frac{3\beta^f_2m^2}{\alpha_f^2}\Big(1+c^2\sum_{k=1}^{N-1}\alpha_k^2\Big)\,.
\eeqn
 we find that $\mu_N^2=\frac{2}{3}\tilde\Lambda$. Hence $\delta M_{\mu\nu}^N$ is a partially massless spin-2 field whose equations are invariant under the gauge transformations~(\ref{PMtrafo}).

\subsection{Perturbative Weyl invariance}
In this section we try to generalise the bimetric analysis of Ref.~\cite{Hassan:2013pca, Hassan:2015tba} to multimetric theory with centre coupling. In bimetric theory it is possible to derive two sets of effective equations, each involving only one of the metrics, as an infinite expansion in curvatures/derivatives. It then turns out that the unique model with scale invariant proportional backgrounds leads to effective equations which start only at fourth order in derivatives. For instance, the equations involving only $\gmn$ assume the form,
\beqn
B_{\mu\nu}(g)+\mathcal{O}\left(R^3\right)=0\,.
\eeqn
Here $B_{\mu\nu}$ is a quantity quadratic in curvatures, known as the Bach tensor,\footnote{The explicit expression for the Bach tensor is provided in~(\ref{Bachdef}). It appears in the equation of motion of conformal gravity, which reads $B_{\mu\nu}=0$~\cite{Bach, Stelle:1976gc, Fradkin:1981iu}.} which under local Weyl transformations of the metric, $\gmn\mapsto\phi(x)\gmn$ with gauge parameter $\phi$, transforms covariantly, $B_{\mu\nu}\mapsto\phi^{-1}B_{\mu\nu}$. Hence, the equations are invariant under these transformations in their lowest order. Terms with more that four derivatives have been neglected. It turns out that the (infinitesimal) Weyl transformations of the two metrics can be extended perturbatively by higher-derivative terms and leave the effective equations invariant at least up to sixth order in derivatives~\cite{Hassan:2015tba}. Whether this construction works for all orders and can be summed up into a closed expression with finite number of derivatives is still an open question.

In multimetric theory, it is in general not obvious how to obtain such effective higher-derivative equations involving only one metric. The best we can do is to treat the interactions as independent bimetric couplings and look for perturbative Weyl invariance under this restrictive assumption.

\subsubsection{Three metrics}\label{sec:pertWeyl3}
Once more we begin with the illustrative example of three metrics $\gmn$, $\fmn$ and $\hmn$. Recall that their equations of motion are of the form,
\begin{subequations}\label{eomthree}
\beqn
\mathcal{G}_{\mu\nu}(g)+m^2\Big[V_{\mu\nu}(g,f;\beta_n^{f})+V_{\mu\nu}(g,h;\beta_n^{h})\Big]&=&0\,,\label{geqtm}
\\
\mathcal{G}_{\mu\nu}(f)+\frac{m^2}{\alpha_f^2}V_{\mu\nu}(f, g;\beta^{f}_{4-n})=0\,, \qquad
\mathcal{G}_{\mu\nu}(h)+\frac{m^2}{\alpha_h^2}V_{\mu\nu}(h, g;\beta^{h}_{4-n})&=&0\,,\label{fheq} 
\eeqn
\end{subequations}
The idea is now to obtain solutions for $\hmn$ and $\fmn$ from the $\gmn$ equations (\ref{geqtm}) as well as two different expressions for $\gmn$ from the $\fmn$ and $\hmn$ equations~(\ref{fheq}), respectively. 
In other words, we solve the equations for the metrics that do not appear with derivatives, such that all manipulations are algebraical. The solutions obtained in this way will be infinite perturbative expansions in curvature terms. 

It is important to note that these manipulations are not equivalent to integrating out the metrics because we are solving the wrong equations for $\hmn$ and $\fmn$. It is therefore not entirely clear how to interpret the resulting effective equations of motion for the metrics derived below. In particular, our method only works at the level of equations of motion but would not result in an equivalent action. The difference to the usual procedure of eliminating fields from the action is discussed in detail in Ref.~\cite{Hassan:2013pca}. Addressing this issue in full detail lies beyond the scope of the present manuscript.

\paragraph{Solving for $\gmn$:}
Let us explain how the procedure works in the $\fmn$ equations; the manipulations in the $\hmn$ equations will be entirely analogous. The details of our calculation can be found in appendix~\ref{app:pertfh}; here we just give a qualitative overview of the procedure.
We start by making the following perturbative ansatz for the square-root matrix $S^{-1}=\sqrt{f^{-1}g}$,
\begin{align}\label{ansatzs1}
{(S^{-1})^\mu}_\nu=b\delta^\mu_\nu&+\tfrac{1}{\mu_f^2}\Big(b_1{P^\mu}_\nu+b_2\Tr\,P\,\delta^\mu_\nu\Big) \nn\\
&+ \tfrac{1}{\mu_f^4}\Big( d_1{(P^2)^\mu}_\nu+d_2\Tr\,P\,{P^\mu}_\nu  
+d_3\Tr\,P^2\delta^\mu_\nu+d_4(\Tr\,P)^2\delta^\mu_\nu \Big) 
+\mathcal{O}\left(\tfrac{P^3}{\mu_f^6}\right)\,,
\end{align}
All terms on the right hand side are curvatures of the metric $\fmn$, which we have also used to raise indices and take traces. 
For later purpose, we have formulated the curvature expansion in terms of the Schouten tensor,
\beqn
P_{\mu\nu}(f)=R_{\mu\nu}(f)-\tfrac{1}{6}\fmn R(f)\,.
\eeqn
The ansatz contains arbitrary (in general complex) coefficients $b, b_i, d_i,\hdots$ which will be determined when we plug it into the equations of motion. The mass scale suppressing the higher-order terms is $\mu_f=m/\alpha_f$. 

Now we use (\ref{ansatzs1}) to replace all occurrences of $\gmn$ in the $\fmn$ equations and require the coefficients of each type of term to vanish separately. This procedure fixes all coefficients in our ansatz. The resulting expression for $S^{-1}$ can be squared to yield an expression for $\gmn=f_{\mu\rho}(S^{-2})^\rho_{~\nu}$. The latter turns out to have the following structure,
\begin{align}\label{gsolP}
\gmn=b^2\fmn+x_gP_{\mu\nu}+y_gP_{\mu\rho}{P^\rho}_\nu
+z_g\Big[\tfrac{1}{3}e_2( P)\delta^\mu_{~\nu}-\Tr\,P\,P_{\mu\nu}\Big]
+\mathcal{O}\left(\tfrac{P^3}{\mu^6}\right)\,.
\end{align}
The exact expressions for the coefficients $x_g$, $y_g$ and $z_g$, which are now fixed in terms of parameters in the trimetric action, can be read off from equation~(\ref{sols}) in the appendix.

We can obtain an analogous solution for $\gmn$ from the $\hmn$ equation and it will have the exact same form as~(\ref{gsolP}) but with curvatures in terms of $\hmn$, suppressing mass scale $m/\alpha_h$ and interaction parameters $\beta_n^h$ instead of $\beta_n^f$. Up until now we have thus derived two expressions for the central metric of the form $\gmn=\gmn(f; \alpha_f, \beta_n^f)$ and $\gmn=\gmn(h,\alpha_h, \beta_n^h)$.

\paragraph{Solving for $\fmn$ and $\hmn$:}
When trying to repeat the above steps in order to obtain perturbative solutions for the two satellite metrics, we face an obstruction: The $\gmn$ equations~(\ref{geqtm}) contain potential terms for both $\fmn$ and $\hmn$ and it is not possible to obtain a general perturbative solution for either of the two. At this point, we will therefore make a rather restrictive assumption on the type of solutions to consider.
Namely, we split up~(\ref{geqtm}) in two separate equations and require them both to be satisfied simultaneously. Essentially, we decide to let one fraction of the Einstein tensor $\mathcal{G}_{\mu\nu}(g)$ enter the perturbative solution for $\fmn$ while the other part should belong to $\hmn$. The same is done for the $\beta_0^f$ terms (recall that we can set $\beta_0^h=0$ without loss of generality), which also depend on $\gmn$ alone and can therefore enter either of the two perturbative solutions.
Thus we consider the following splits of these terms in the $\gmn$ equations,
\beqn\label{split}
\mathcal{G}_{\mu\nu}(g)=A\mathcal{G}_{\mu\nu}(g)+(1-A)\mathcal{G}_{\mu\nu}(g)\,,\qquad
\beta_0^{f}=B\beta_0^{f}+(1-B)\beta_0^{f}\,,
\eeqn
with constants $0<A<1$ and $0<B<1$.
Now we use only the first parts $A\mathcal{G}_{\mu\nu}(g)$ and $B\beta_0^{f}$ in obtaining the perturbative solution for $\fmn$ and the remaining terms for $\hmn$. Thus we are essentially working with two separate sets of equations of the form,
\begin{subequations}
\beqn
A\mathcal{G}_{\mu\nu}(g)+Bm^2\beta^f_0\gmn+\left.m^2V_{\mu\nu}(g,f;\beta_n^{f})\right|_{\beta^f_0=0}&=&0\,,\label{splitgeqf}\\
(1-A)\mathcal{G}_{\mu\nu}(g)+(1-B)m^2\beta^f_0\gmn+\left.m^2V_{\mu\nu}(g,h;\beta_n^{h})\right|_{\beta^h_0=0}&=&0\,.\label{splitgeqh}
\eeqn
\end{subequations}
The restriction $0<A<1$ ensures that the sign in front of the kinetic term is maintained in both of these.\footnote{Relaxing the assumptions $0<A<1$ and $0<B<1$ would mean that we did not simply split the terms with pure $\gmn$ dependence but would enable the solutions for $\fmn$ to create extra terms which would then have to be cancelled by terms coming from $\hmn$. In this case we should also allow for such cancellations among all other terms in the perturbative series for the satellite metrics. This would introduce many more free parameters and no definite statements could be made in the following. We therefore exclude this possibility here and restrict ourselves to simple separations of the coupling into two bimetric parts, which both contain the Einstein tensor with the usual positive sign in front.}   
By assuming two separate expansions, we focus on solutions for which the couplings between $(\gmn,\fmn)$ and $(\gmn,\hmn)$ behave like two separate bimetric interactions. Without this strong restriction it is not possible to arrive at the results below.

Following the exact same steps that lead us to the above solution for $\gmn$, we then use (\ref{splitgeqf}) to derive an expression for $\fmn$ in terms of curvatures for~$\gmn$,
\begin{align}\label{fsolP1}
\fmn=a^2\gmn+x_fP_{\mu\nu}+y_f{(P^2)^\mu}_\nu
+z_f\Big[\tfrac{1}{3}e_2( P)\delta^\mu_\nu-\Tr\,P\,{P^\mu}_\nu\Big]
+\mathcal{O}\left(\tfrac{P^3}{\mu^6}\right)\,.
\end{align}
Here, the suppressing scale is $\mu^2=m^2/A$ and the coefficients can be read off from equation~(\ref{fsolP}) in the appendix.
We have thus arrived at a perturbative expression of the form $\fmn=\fmn(g, \beta^f_n, A, B)$ and, in a completely analogous way, we obtain a solution of the form $\hmn=\hmn(g, \beta^h_n, A, B)$ from the remaining part of the $\gmn$ equation.

\paragraph{Effective equations for $\gmn$:}
The next step is to use the perturbative expression~(\ref{fsolP1}) and its analogue to eliminate $\fmn$ and $\hmn$ from their own equation of motion~(\ref{fheq}). We shall discuss this for the $\fmn$ equation, the $\hmn$ case is completely analogous. The detailed calculation and the exact expressions can be found in appendix~\ref{app:pertg}, here we simply state the qualitative result.
Using (\ref{fsolP1}) to replace all occurrences of $\fmn$ in its own equation of motion, we arrive at the following effective equation for $\gmn$,
\begin{align}\label{fEq2}
\Lambda_g\,\gmn+
r_1\mathcal{G}_{\mu\nu}
&+r_2P_{\mu\nu} 
+r_3B_{\mu\nu}
\nn\\
&
+r_4\Big[(s_1+2s_2)P_\mu^{~\rho}
P_{\rho\nu} -2s_2PP_{\mu\nu}-\tfrac{s_2}{3}\gmn\left(P_{\rho\sigma}
P^{\rho\sigma}-P^2\right)\Big]
\nn\\
&-r_5\left[3P P_{\mu\nu}
-2P_\mu^{~\rho} P_{\rho\nu} -\frac{1}{2}g_{\mu\nu}(P^2-P^{\alpha\beta}
P_{\alpha\beta})\right]+\mathcal{O}\left(\tfrac{P^3}{\mu^6}\right) 
=0\,.
\end{align}
Here, all expressions are with respect to $\gmn$ and we have collected some four-derivative terms into the Bach tensor $B_{\mu\nu}$ which is defined in~(\ref{Bachdef}). The $\hmn$ equations will give analogous results with parameters that we shall refer to as $\tilde{\Lambda}_g$, $\tilde{r}_i$ and $\tilde{s}_i$ below.

\paragraph{Effective equations for $\fmn$ and $\hmn$:}
Now we would also like to turn the $\gmn$ equations~(\ref{geqtm}) into effective equations for $\fmn$ and $\hmn$.
Since we have two different expressions for $\gmn$, one in terms of $\fmn$ and one in terms of $\hmn$, this procedure is ambiguous. In the potential contributions to (\ref{geqtm}), it makes sense to replace the central metric by $\gmn(f; \alpha_f, \beta_n^f)$ in $V_{\mu\nu}(g,f;\beta_n^{(f)})$ and by $\gmn(h; \alpha_h, \beta_n^h)$ in $V_{\mu\nu}(g,h;\beta_n^{(h)})$. This results in two separate expressions involving only $\fmn$ and only $\hmn$, respectively, and any other replacement would introduce nonlinear mixings between the two satellite metrics. The ambiguity arises in the kinetic term $\mathcal{G}_{\mu\nu}(g)$ and also in the $\beta_0^f$ terms, which both only involve $\gmn$. Hence we again allow for an arbitrary split,
\beqn\label{split2}
\mathcal{G}_{\mu\nu}(g)=A'\mathcal{G}_{\mu\nu}(g)+(1-A')\mathcal{G}_{\mu\nu}(g)\,,\qquad
\beta_0^{f}=B'\beta_0^{f}+(1-B')\beta_0^{f}\,,
\eeqn
with constants $A'$ and $B'$ which are not necessarily the same as in~(\ref{split}). Now we use $\gmn(f; \alpha_f, \beta_n^f)$ to replace $\gmn$ in $A'\mathcal{G}_{\mu\nu}(g)$ and $B'\beta_0^{f}\gmn$. The effective equations that we derive in this way are thus not unique for fixed interaction parameters but will depend on our choice for $A'$ and $B'$.
The details of this computation are provided in appendix~\ref{app:pertfh}. 
Here we simply write down the resulting terms involving curvatures of $\fmn$ after $\gmn$ has been eliminated from its equations~(\ref{geqtm}) following the above procedure. They have exactly the same structure as in~(\ref{fEq2}) but contain different coefficients,
\begin{align}\label{fEqbla}
&\Lambda_f\,\fmn
+t_1\mathcal{G}_{\mu\nu}(f)
+t_2P_{\mu\nu} 
+t_3B_{\mu\nu}
\nn\\
&
+t_4\Big[(\sigma_1+2\sigma_2)P_\mu^{~\rho}
P_{\rho\nu} -2\sigma_2PP_{\mu\nu}-\tfrac{\sigma_2}{3}\fmn\left(P_{\rho\sigma}
P^{\rho\sigma}-P^2\right)\Big]
\nn\\
&-t_5\left[3P P_{\mu\nu}
-2P_\mu^{~\rho} P_{\rho\nu} -\frac{1}{2}f_{\mu\nu}(P^2-P^{\alpha\beta}
P_{\alpha\beta})\right]
+\mathcal{O}\left(\tfrac{P^3}{\mu_f^6}\right) ~+~(\hmn\,\text{terms})
=0\,.
\end{align}
In the remaining terms of the $\gmn$ equations, we replace $\gmn$ by $\gmn(h; \alpha_h, \beta_n^h)$ and obtain an entirely analogous expression in terms of $\hmn$ and its curvatures. It has the same form as the $\fmn$ terms in~(\ref{fEqbla}), containing its own coefficients which we shall refer to as $\Lambda_h$, $\tilde{t}_i$ and $\tilde{\sigma}_i$ below.

\paragraph{Weyl invariance:}

We would now like to find the parameter values for which the effective equations, (\ref{fEq2}) and (\ref{fEqbla}), exhibit Weyl invariance to lowest order in the expansion. The details of this calculation can be found in appendix~(\ref{app:weyl}).
Since neither the cosmological constant term proportional to $\gmn$ nor any linear curvature terms proportional to $R_{\mu\nu}$ and $\gmn R$ can be invariant under Weyl transformations, these contributions have to vanish. We therefore demand the following constraints on the expansion coefficients,
\beqn\label{constec}
\Lambda_g=r_1=r_2&=&0\,,\qquad
\tilde{\Lambda}_g=\tilde{r}_1=\tilde{r}_2=0\,,\nn\\
\Lambda_f=t_1=t_2&=&0\,,\qquad
\Lambda_h=\tilde{t}_1=\tilde{t}_2=0\,.
\eeqn
From the explicit expressions in the appendix, it is obvious that these requirements automatically imply,
\beqn
r_4=r_5=0\,,\qquad
\tilde{r}_4=\tilde{r}_5=0\,,\qquad
t_4=t_5=0\,,\qquad
\tilde{t}_4=\tilde{t}_5=0\,.
\eeqn 
Thus all terms in the effective equations except for the Bach tensor contributions vanish and we are left with,
\begin{subequations}  
\beqn
r_3 B_{\mu\nu}(g)+\mathcal{O}\Big(\tfrac{P^3(g)}{\mu^6}\Big)&=&0\,,\label{bachg}\\
t_3 B_{\mu\nu}(f)+\tilde{t}_3 B_{\mu\nu}(h)+\mathcal{O}\Big(\tfrac{P^3(f)}{\mu_f^6}\Big)+\mathcal{O}\Big(\tfrac{P^3(h)}{\mu_h^6}\Big)&=&0
\eeqn
\end{subequations}
 Recall that in fact we obtain two copies of (\ref{bachg}), one from the $\fmn$ and one from the $\hmn$ equations. To lowest order in derivatives, these equations are invariant under Weyl transformations of all three metrics.
 
The requirements on the expansion coefficients in~(\ref{constec}) translate into constraints on the interaction parameters.
As discussed in detail in the appendix, they can be shown to be equivalent to,
\begin{subequations}\label{betaweyl}
\beqn
\beta^f_1=\beta^f_3=\beta^f_1=\beta^f_3&=&0\,,\\
\alpha_f^2\beta_4^h\beta_0^f+9\alpha_h^2\beta_2^h\beta_2^f-9\alpha_f^2(\beta_2^h)^2-3\beta_2^f\beta_4^h&=&0\,,\\
3\alpha_f^2\beta_2^f\beta_4^h+3\alpha_h^2\beta_2^h\beta_4^f-\beta_4^f\beta_4^h&=&0\,.
\eeqn
\end{subequations}
They also fix the splits of the $\gmn$ equations in (\ref{split}) and (\ref{split2}) in the following way,
\beqn
A=A'=\tfrac{3\alpha_f^2\beta_2^f}{\beta_4^f}\,,\qquad
B=B'=\tfrac{9(\beta_2^f)^2}{\beta_0^f\beta_4^f}\,.
\eeqn 
We have therefore shown that, treating the trimetric coupling as two separate bimetric interactions, we can arrive at effective equations with Weyl invariance at lowest order if and only if the interaction parameters satisfy~(\ref{betaweyl}). 
Remarkably, these conditions are identical to the constraints~(\ref{beta13}) and~(\ref{betaprop}) which we found in demanding scale invariance of the proportional backgrounds.
This result suggests an intimate relation between scale invariance and perturbative Weyl symmetry in trimetric theory.

\subsubsection{Centre coupling}

Some of the trimetric results can directly be generalised to the case $N$ fields $\fmn^i$ coupled to a central metric $\gmn$. Deriving the most general constraints on the parameters requires larger computational effort which is beyond the scope of our present work. We therefore restrict ourselves to the simplest solution which corresponds to an action with interchange symmetry among the satellite metrics $\fmn^i$. 

We solve each of the $N$ equations of motions~(\ref{eomsat}) of the satellite metrics perturbatively for $\gmn$. These expressions are inserted into the equations~(\ref{eomcen}) of $\gmn$, in which we split the Einstein tensor and $\beta_0^1$ term as,
\beqn
\mathcal{G}_{\mu\nu}(g)&=& \frac{1}{\sum_{k=1}^N\alpha_k^2} \Big(\alpha_1^2\mathcal{G}_{\mu\nu}(g)+\alpha_2^2\mathcal{G}_{\mu\nu}(g)+\hdots \alpha_N^2\mathcal{G}_{\mu\nu}(g)\Big)\,,\\
\beta_0^1\gmn&=& \frac{1}{\sum_{k=1}^N\alpha_k^2} \Big(\alpha_1^2\beta_0^1\gmn+\alpha_2^2\beta_0^1\gmn+\hdots \alpha_N^2\beta_0^1\gmn\Big)\,.
\eeqn
In the $i$th term as well as in the potential terms involving $\fmn^i$ we use the perturbative solution coming from the $\fmn^i$ equations to eliminate $\gmn$. Then we use the same split in order to obtain perturbative solutions for each $\fmn^i$. Again it should be noted that this is a very restrictive assumption and does not yield the most general solutions to the equations. Nevertheless, this method will result in a set of effective perturbative equations. Their lowest orders assume the form, 
\begin{subequations}  
\beqn
 B_{\mu\nu}(g)+\mathcal{O}\Big(\tfrac{P^3(g)}{\mu^6}\Big)&=&0\,,\label{bachgc}\\
t_1B_{\mu\nu}(f^1)+ t_2B_{\mu\nu}(f^2)+\hdots+ t_NB_{\mu\nu}(f^N)+\sum_{i=1}^N\mathcal{O}\Big(\tfrac{P(f_i)^3}{\mu_i^6}\Big)&=&0\,,
\eeqn
\end{subequations}
if and only if the interaction parameters satisfy
\beqn
\beta_n^{(i)}&=&\frac{\alpha_i^2}{\alpha_1^2}\beta_n^{(1)}\,,\qquad n=1,2,3,4\,,~~~i=2,\hdots N\,,\nn\\
\alpha_1^2\beta^{(1)}_0&=&3\beta_2^{(1)}\,,\qquad 
3\beta^{(1)}_2\sum_{k=1}^{N}\alpha_k^2=\beta_4^{(1)}\qquad 
\beta^{(1)}_1=\beta^{(1)}_3=0\,.
\eeqn
These are exactly the same coefficients as in~(\ref{PMbetas1}) and~(\ref{centcoupbeta}), which again relates the scale invariance of the proportional backgrounds to the emergence of Weyl symmetry at the perturbative level. Of course we have been not completely general in this analysis for $N$ metrics and the result merely demonstrates that there exists a (very symmetric) parameter choice that satisfies both requirements. Nevertheless, from our point of view it seems quite likely that their equivalence is realised for any number of metrics.

\section{Discussion}\label{sec:discussion}

In this work we have analysed generalisations of ghost-free bimetric theory describing interactions for multiple tensor fields. Our main focus has been on $N$ fields coupled to one central metric. For these models we have obtained the spectrum of mass eigenstates in terms of roots to an $N$th order polynomial equation. Under some additional symmetry assumptions, the expressions for the mass eigenstates and eigenvalues assume a simple form which is a direct generalisation of the bimetric case. An example for this simplification is a parameter choice for which the multimetric action becomes symmetric in the interchange of any two satellite metrics.

For $N=2$ we have explicitly derived the complete parameter region for which the proportional background solutions are invariant under constant scale transformations. We have then shown that for the exact same parameter values it is possible to obtain perturbative solutions to the equations of motion whose lowest orders are invariant under Weyl transformations of the metrics. Let us comment on this outcome in a bit more detail.

In bimetric theory a similar result has been obtained and the corresponding parameter constraints read~\cite{Hassan:2012gz, Hassan:2013pca},
\beqn\label{bipm2}
\beta_1=\beta_3=0\,,\qquad
\alpha^4\beta_0=3\alpha^2\beta_2=\beta_4\,.
\eeqn
In this case the equivalent requirements of scale or Weyl invariance fix all interaction parameters in terms of one overall scale and the resulting model is symmetric in the interchange of the two metrics. In contrast, the corresponding multimetric model specified by (\ref{betaweyl}) is not unique but still contains two free parameters plus the overall scale. Only if in addition we demand interchange symmetry among the two satellite metrics, we arrive at the unique set of parameters~(\ref{PMbetas2}). The interactions in this particular model look a bit similar to the bimetric ones but they do not simply correspond to two copies of~(\ref{bipm2}).  
Note however that, even for general~$N$, all models with scale invariant proportional backgrounds share the property $\beta^{(i)}_1=\beta^{(i)}_3=0$ for all~$i$.

In demanding scale invariance of the proportional backgrounds our analysis is completely general. In the derivation of the perturbative solutions to the equations, however, we had to make a restrictive assumption in order to arrive at separate expansions for each metric. The equations for the central metric $\gmn$ contain both satellite metrics and therefore cannot yield a general algebraical solution for one of them in terms of only $\gmn$ and its curvatures. In order to arrive at such an expression, we had to split up the curvature terms for $\gmn$ and distribute them among the two satellite metrics. This means that we have treated the centre coupling as if it consisted of two separate bimetric couplings. The resulting solutions are not at all general but without the restriction we cannot make any statement about the appearance of Weyl invariance.
\\
On the other hand, it is important to note that we are not simply dealing with two entirely independent bimetric couplings. If that were the case, we could have simply used the bimetric results and would have arrived at the Weyl invariant Bach equation in a trivial way. In this case we would have found the bimetric parameters~(\ref{bipm2}) in both sectors. The nontrivial part in our analysis is that the split of the Einstein tensor and $\beta^f_0$ term in the $\gmn$ equations connects the two sets of parameters in the $\fmn$ and the $\hmn$ coupling. As a result, we arrived at parameter values that are different from the bimetric ones and the nontrivial and remarkable result is that they coincide with the parameters obtained by demanding scale invariance on the proportional solutions.

 Our results suggest a deeper connection of scale and Weyl invariance in spin-2 interactions. In general, a Weyl invariant theory is also scale invariant because the local gauge symmetry of course implies the global invariance. The reverse statement however is not automatically true.
In our case, the implications in any direction are not that obvious because we are dealing neither with a fully Weyl invariant nor fully scale invariant action. We merely see these symmetries emerge for some particular solutions to the equations of motion and, on top of that, the Weyl transformations only appear at lowest order in a perturbative expansion and never leave the full solutions invariant. It is therefore not at all clear why both symmetries are found in exactly the same set of trimetric models and it would be interesting to develop a better understanding of the underlying reasons for this connection.

We furthermore saw that, just as in bimetric theory, the models specified by (\ref{betaweyl}) contain a partially massless (PM) spin-2 mode in their mass spectrum. It thus seems that the additional structures are also related (but not equivalent) to the presence of the corresponding gauge symmetry in the linear theory.\footnote{Note that simply imposing the Higuchi bound~(\ref{higuchi}) for one of the massive modes only fixes one interaction parameter and does therefore not uniquely single out the models with additional structures.} At first sight this may not come as a surprise considering that the Weyl invariant theory of conformal gravity with equation of motion $B_{\mu\nu}=0$ also contains a PM field. But in our case the Bach tensor only emerges at lowest order in a perturbative series and the connection to the presence of linear PM fields is not obvious. Moreover, in multimetric theory the massless and the PM field are healthy, while the massive fields correspond to Higuchi ghosts. One may therefore argue that the theories in this respect are on the same footing as conformal gravity which inevitably contains a ghost.
\\
It is possible that the identified models extend part of the PM gauge symmetry beyond the linear theory in a way that is not yet fully understood. This is definitely the case for the proportional backgrounds with parameters~(\ref{betaweyl}) whose scale invariance can be regarded as the constant part of the linear PM transformations~(\ref{PMtrafo}), transferred to the background level~\cite{Hassan:2012gz}. 
At this stage we can only speculate, but in principle PM fields could help in addressing the cosmological constant problem~\cite{Weinberg:1988cp}.
This is due to the fact that partial masslessness relates the background curvature to the spin-2 mass whose value is protected against receiving large quantum corrections by the diffeomorphism invariance of the massless theory. It is therefore possible that, in a nonlinear  theory with enhanced symmetry, also the cosmological constant could assume a naturally small value. 
However, as already stated in the beginning of section~\ref{sec:scinv}, our setup involving only spin-2 degrees of freedom most likely cannot describe the nonlinear interactions of PM fields. Nevertheless, nonlinear theories involving PM fluctuations around de Sitter solutions are therefore an interesting subject that deserves further exploration.

In summary, our work revealed that certain properties of bimetric theory carry over to the multimetric case and it is therefore unlikely that these are simply structural coincidences. As a result, we believe that it is worth studying this issue further in order to develop a full understanding of the additional structures and symmetries present in multiple spin-2 interactions.

\vspace{0.5cm}

{\bf Acknowledgment}: The authors would like to thank Niklas Beisert, Fawad Hassan and Mikael von Strauss for fruitful discussions.
ASM acknowledges funding from the ERC grant no.~615203 under the FP7 and the Swiss National Science Foundation through the NCCR SwissMAP.

\appendix

\section{Perturbative expansion of equations}\label{app:pert}

This appendix contains the derivation of the effective equations and constraints on the interaction parameters in the trimetric case, as discussed in section~\ref{sec:pertWeyl3}.

\subsection{The effective equations for $\fmn$ and $\hmn$}\label{app:pertfh}

Using the ansatz~(\ref{ansatzs1}) in the equations of motion for $\fmn$, the lowest orders in the solution for $S^{-1}$ are determined to be of the following form,
\begin{align}\label{sols}
{(S^{-1})^\mu}_\nu=b\delta^\mu_{~\nu}+\tfrac{b}{\sigma_1\mu_f^2}{P^\mu}_\nu +\tfrac{b\sigma_2}{(\sigma_1)^3\mu_f^4}\Big[{(P^2)^\mu}_\nu-\Tr\,P\,{P^\mu}_\nu+\tfrac{1}{3}e_2( P)\delta^\mu_{~\nu}\Big] 
+\mathcal{O}\left(\tfrac{P^3}{\mu_f^6}\right)\,,
\end{align}
where all expressions on the right-hand side are in terms of $\fmn$. The suppressing mass scale is $\mu_f=m/\alpha_f$, the expansion  coefficients are given in terms of linear combinations of the $\beta^f_n$ parameters,
\beqn
\sigma_n&\equiv& \sum_{k=n}^{3}{3-n\choose k-n}b^k\beta^f_{n-k}\,,\qquad n=1,2,3\,,\\
\sigma_0&\equiv&\beta_4^f+3b\beta_3^f+3b^2\beta_2^f+b^3\beta_1^f\,,
\eeqn
and $b$ is the solution to the polynomial equation $\sigma_0=0$.
We can then use $\gmn=f_{\mu\rho}{(S^{-2})^\rho}_\nu$ to arrive at the perturbative solution for $\gmn$,
\begin{align}\label{fsolP2}
\gmn=b^2\fmn+\tfrac{2b^2}{\sigma_1\mu_f^2}P_{\mu\nu}+\tfrac{b^2(\sigma_1+2\sigma_2)}{\sigma_1^3\mu_f^4}P_{\mu\rho}P^\rho_{~\nu}
+\tfrac{2b^2\sigma_2}{\sigma_1^3\mu_f^4}\Big[\tfrac{1}{3}e_2( P)\fmn&-\Tr\,P\,P_{\mu\nu}\Big]\nn\\
&+\mathcal{O}\left(\tfrac{P^3}{\mu_f^6}\right)\,.
\end{align}
Next we would like to use this expression (and the analogous one derived from the $\hmn$ equation) to eliminate $\gmn$ from its own equation of motion. 
As outlined in section~\ref{sec:pertWeyl3}, we split up the Einstein tensor in the $\gmn$ equations in the following way,
\beqn
\mathcal{G}_{\mu\nu}(g)=A'\mathcal{G}_{\mu\nu}(g)+(1-A')\mathcal{G}_{\mu\nu}(g)\,,
\eeqn
with $0<A'<1$, and replace $\gmn$ by~(\ref{fsolP2}) in the first term.
This gives, 
\beqn
A'{\mathcal{G}}_{\mu\nu}(g)&=&A'{\mathcal{G}}_{\mu\nu}(f)-\tfrac{A'}{\sigma_1\mu_f^2}
\Big( \nabla^2P_{\mu\nu}+\nabla_\mu\nabla_\nu 
P-\nabla^\rho\nabla_\mu P_{\rho\nu}-\nabla^\rho\nabla_\nu P_{\rho\mu} \nn\\
&~&\hspace{100pt} 
+~3PP_{\mu\nu}-\fmn\left[P^{\alpha\beta}P_{\alpha\beta}+\tfrac{1}{2}P^2
  \right] \Big)+\mathcal{O}\left(\tfrac{P^3}{\mu_f^4}\right)\nn\\
 &=&A'{\mathcal{G}}_{\mu\nu}(g)+\tfrac{A'}{\sigma_1\mu_f^2}B_{\mu\nu} \nn\\
&~&\hspace{20pt} 
 +~\tfrac{A'}{\sigma_1\mu_f^2}\Big(2P_{\mu\rho}P^\rho_{~\nu}-3PP_{\mu\nu}+\tfrac{1}{2}\fmn\left[P^{\alpha\beta}P_{\alpha\beta}+P^2
  \right] \Big)+\mathcal{O}\left(\tfrac{P^3}{\mu_f^4}\right)\,.
\eeqn
Here we have collected some of the terms with four derivative into the Bach tensor,
\beqn\label{Bachdef}
   B_{\mu\nu}\equiv 
   -\nabla^2 P_{\mu\nu}
   -\nabla_{\mu}\nabla_{\nu}P^\rho_{~\rho}
   +\nabla^{\rho}\nabla_{\mu}P_{\nu\rho}
   +\nabla^{\rho}\nabla_{\nu}P_{\mu\rho}
   -2P_{\mu\rho}P^\rho_{~\nu}
   +\tfrac{1}{2}\fmn P^{\rho\sigma}P_{\rho\sigma}
   \,.
\eeqn 
Recall that in the potential contributions we can set $\beta_0^h=0$ in order to remove its degeneracy with $\beta_0^f$. The corresponding term which enters only the $\gmn$ equations is then of the form $\beta_0^f\gmn$.
For this term, which also depends on $\gmn$ alone, we do a similar split,
\beqn
\beta_0^{f}=B'\beta_0^{f}+(1-B')\beta_0^{f}\,,
\eeqn
with $0<B'<1$ and use only the first term for the replacement in terms of $\fmn$ curvatures.
The contributions from the interaction potential to the $\gmn$ equation evaluated on~(\ref{fsolP2}) then become, 
\begin{align}
m^2{V}_{\mu\nu}(g,f;\beta_n^f)&=\tfrac{m^2\Omega}{b^2} 
\fmn+\tfrac{\alpha_f^2}{b^2}\mathcal{G}_{\mu\nu}(f)+
\tfrac{2\alpha_f^2\Omega}{b^2\sigma_1}P_{\mu\nu}\nn\\
&+\tfrac{\alpha_f^2}{b^2\sigma_1^3\mu_f^2}\left[\kappa_1P^\rho_\mu P_{\rho\nu}+
\kappa_2 PP_{\mu\nu}+\tfrac{1}{6}\fmn(\kappa_3P^{\alpha\beta}P_{\alpha\beta}-\kappa_2P^2)
\right]+\mathcal{O}\left(\tfrac{P^3}{\mu_f^4}\right)\,.
\end{align}
The expansion coefficients are given by,
\begin{align}
	\kappa_1=2\sigma_1^2+\Omega'(\sigma_1+2\sigma_2)\,,\qquad
	\kappa_2=-3\sigma_1^2-2\sigma_2\Omega'\,,\qquad
	\kappa_3=3\sigma_1^2-2\sigma_2\Omega'\,,
\end{align}
where we have defined,
\begin{align}
	\Omega'=b\beta^f_3+3b^2\beta^f_2+3b^3\beta^f_1+B'b^4\beta^f_0\,.
\end{align}
Combining the kinetic and potential terms, we can write the first part of the $\gmn$ equation of motion
as a higher-derivative equation for $\fmn$, 
\begin{align}\label{efffmnhmn}
&\tfrac{\Omega'}{b^2}\,\fmn+\tfrac{1}{\mu_f^2}
\Big[\tfrac{1}{b^2}+\tfrac{1}{\alpha^2_f}A'\Big]\mathcal{G}_{\mu\nu}(f)
+\tfrac{2\Omega'}{b^2\sigma_1\mu_f^2}\,P_{\mu\nu} 
\nn\\
&+A'\tfrac{\alpha_f^2}{\mu_f^4 \sigma_1}B_{\mu\nu}
+\tfrac{\alpha^2_f\Omega'}{b^2\sigma_1^3\mu_f^4}\Big[(\sigma_1+2\sigma_2)P_\mu^{~\rho}
P_{\rho\nu} -2\sigma_2PP_{\mu\nu}-\tfrac{\sigma_2}{3}\fmn\left(P_{\rho\sigma}
P^{\rho\sigma}-P^2\right)\Big]
\nn\\
&-\tfrac{1}{\sigma_1\mu_f^4}\Big[\tfrac{\alpha^2_f}{b^2}+A'\Big]\left[3P P_{\mu\nu}
-2P_\mu^{~\rho} P_{\rho\nu} -\frac{1}{2}f_{\mu\nu}(P^2-P^{\alpha\beta}
P_{\alpha\beta})\right]\nn\\
& 
+\mathcal{O}\left(\tfrac{P^3}{\mu_f^6}\right) 
+\big(\hmn\,\text{terms}\big)
~=~0\,,
\end{align}
The $\hmn$ terms, obtained from solving the $\hmn$ equations perturbatively for $\gmn$ and eliminating $\gmn$ in the remaining terms of its equations, have the exact same structure.

\subsection{The effective equations for $\gmn$}\label{app:pertg}

We start by deriving perturbative solutions $\fmn$ and $\hmn$ from the $\gmn$ equations.
First we have to split the ambiguous terms in the following general way,
\beqn
\mathcal{G}_{\mu\nu}(g)=A\mathcal{G}_{\mu\nu}(g)+(1-A)\mathcal{G}_{\mu\nu}(g)\,,\qquad
\beta_0^{f}=B\beta_0^{f}+(1-B)\beta_0^{f}\,,
\eeqn
with $0<A<1$ and $0<B<1$. Then we treat the equations as if they were two decoupled bimetric interactions.
In terms of $\mu^2=m^2/A$, we obtain a perturbative solution for $S=\sqrt{g^{-1}f}$ of the form,
\begin{align}\label{ansatzs}
{S^\mu}_\nu=a\delta^\mu_{~\nu}+\tfrac{a}{s_1\mu^2}{P^\mu}_\nu +\tfrac{as_2}{s_1^3\mu^4}\Big[{(P^2)^\mu}_\nu-\Tr\,P\,{P^\mu}_\nu+\tfrac{1}{3}e_2( P)\delta^\mu_{~\nu}\Big] 
+\mathcal{O}\left(\tfrac{P^3}{\mu}\right)\,.
\end{align}
Here the suppressing mass scale is $\mu=m/A$, we have introduced a new set of linear combinations of the $\beta_n$ parameters,
\beqn
s_n&\equiv& \sum_{k=n}^{3}{3-n\choose k-n}a^k\beta^f_k\,,\qquad n=1,2,3\,,\\
s_0&\equiv&B\beta^f_0+3a\beta_1^f+3a^2\beta_2^f+a^3\beta_3^f\,,
\eeqn
and $a$ is the solution to the polynomial equation $s_0=0$.
We can then use $\fmn=g_{\mu\rho}{(S^2)^\rho}_\nu$ to arrive at the perturbative solution for $\fmn$,
\begin{align}\label{fsolP}
\fmn=a^2\gmn+\tfrac{2a^2}{s_1\mu^2}P_{\mu\nu}+\tfrac{a^2(s_1+2s_2)}{s_1^3\mu^4}{(P^2)^\mu}_\nu
+\tfrac{2a^2s_2}{s_1^3\mu^4}\Big[\tfrac{1}{3}e_2( P)\delta^\mu_\nu&-\Tr\,P\,{P^\mu}_\nu\Big]\nn\\
&+\mathcal{O}\left(\tfrac{P^3}{\mu^6}\right)\,.
\end{align}
Next we use this expression to eliminate $\fmn$ from its own equations of motion. The computation is completely analogous to the one in the previous subsection and we arrive at the following effective higher-derivative equations for $\gmn$,
\begin{align}\label{fEq2nd}
&\tfrac{\Omega}{a^2\alpha^2}\,\gmn+\tfrac{1}{\mu^2}
\Big[\tfrac{1}{A}+\tfrac{1}{a^2\alpha^2}\Big]\mathcal{G}_{\mu\nu}
+\tfrac{2\Omega}{a^2\alpha^2s_1\mu^2}\,P_{\mu\nu} 
\nn\\
&+\tfrac{1}{\mu^4 s_1}B_{\mu\nu}
+\tfrac{\Omega}{a^2\alpha^2s_1^3\mu^4}\Big[(s_1+2s_2)P_\mu^{~\rho}
P_{\rho\nu} -2s_2PP_{\mu\nu}-\tfrac{s_2}{3}\gmn\left(P_{\rho\sigma}
P^{\rho\sigma}-P^2\right)\Big]
\nn\\
&-\tfrac{1}{s_1\mu^4}\Big[\tfrac{1}{A}+\tfrac{1}{a^2\alpha^2}\Big]\left[3P P_{\mu\nu}
-2P_\mu^{~\rho} P_{\rho\nu} -\frac{1}{2}g_{\mu\nu}(P^2-P^{\alpha\beta}
P_{\alpha\beta})\right]+\mathcal{O}\left(\tfrac{P^3}{\mu^6}\right) 
=0\,,
\end{align}
Again we have collected some of the terms with four derivative into the Bach tensor $B_{\mu\nu}$ defined in~(\ref{Bachdef}).
Moreover, we have used the abbreviation,
\begin{align}
	\Omega=a\beta^f_1+3a^2\beta^f_2+3a^3\beta^f_3+a^4\beta^f_4\,.
\end{align}
Using the remaining parts of the $\gmn$ equations to obtain a perturbative solution for $\hmn$, we can derive another set of effective equations for $\gmn$ from the $\hmn$ equations. These have the exact same structure as~(\ref{fEq2nd}).

\subsection{Parameter constraints}\label{app:weyl}
Now we demand the lowest order of the effective equations (\ref{efffmnhmn}) and (\ref{fEq2nd}) as well as its analogue obtained from the $\hmn$ equations to consist of the Bach tensor alone. From~(\ref{fEq2nd}) we get the constraints,
\beqn
a^2=-\frac{A}{\alpha_f^{2}}\,,\qquad
\Omega=0\,.
\eeqn
Since $A>0$, the first of these implies that $a=\pm i{\sqrt{A}}/{\alpha_f}$. Evaluated on this imaginary value for $a$, the second condition becomes a complex equation whose real and imaginary parts give,
\beqn
\alpha_f^2\beta^f_1=3A\beta^f_3\,,\qquad 
3\alpha_f^{2}\beta^f_2=A\beta^f_4\,.
\eeqn
From $s_0=0$, which is the defining equation for $a$, we furthermore get the relations,
\beqn
3\alpha_f^2\beta^f_1=A\beta^f_3\,,\qquad 
\alpha_f^{2}B\beta^f_0=3A\beta_2^f\,.
\eeqn
Together these imply the following four relations,
\beqn\label{fourrel}
\beta^f_1=\beta^f_3=0\,,\qquad
3\alpha_f^{2}\beta^f_2=A\beta^f_4\,,\qquad
B\alpha_f^{2}\beta^f_0=3A\beta_2^f\,.
\eeqn
Requiring the Bach tensor to constitute the lowest order in~(\ref{efffmnhmn}) gives the conditions,
\beqn
b^2=-\frac{\alpha_f^2}{A'}\,,\qquad
\Omega'=0\,.
\eeqn
Using the above result, $\beta_1^f=\beta_3^f=0$, and the first of these conditions, we find that the second reduces to,
\beqn
3A'\beta^f_2=\alpha_f^2B'\beta^f_0\,.
\eeqn
Moreover, the defining equation for $b$ is $\sigma_0=0$ from which it follows that,
\beqn 
3\alpha_f^{2}\beta^f_2=A'\beta^f_4\,.
\eeqn
Comparing the last two expressions with (\ref{fourrel}) we find that the splits of the $\gmn$ equations must be equal,
\beqn
A=A'\,,\qquad
B=B'\,.
\eeqn
Next, we turn to the $\hmn$ terms in~(\ref{efffmnhmn}) for which, since $(1-A) >0$, things work exactly as above. The resulting constraints therefore read,
\beqn
\beta_1^h=\beta_3^h=0\,,\qquad 
3\alpha_h^{2}\beta^h_2=(1-A)\beta^h_4\,,\qquad
3(1-A)\beta^h_2=\alpha_h^2(1-B)\beta^h_0\,.\qquad
\eeqn
These conditions together with (\ref{fourrel}) imply that the $\gmn$ equations must be split according to,
\beqn
A=\tfrac{3\alpha_f^2\beta_2^f}{\beta_4^f}\,,\qquad
B=\tfrac{9(\beta_2^f)^2}{\beta_0^f\beta_4^f}\,.
\eeqn
In addition we finally arrive at the following constraints on the interaction parameters,
\begin{subequations}
\beqn
\beta^f_1=\beta^f_3=\beta^f_1=\beta^f_3&=&0\,,\\
\alpha_f^2\beta_4^h\beta_0^f+9\alpha_h^2\beta_2^h\beta_2^f-9\alpha_f^2(\beta_2^h)^2-3\beta_2^f\beta_4^h&=&0\,,\\
3\alpha_f^2\beta_2^f\beta_4^h+3\alpha_h^2\beta_2^h\beta_4^f-\beta_4^f\beta_4^h&=&0\,,
\eeqn
\end{subequations}
which are exactly identical to those found in section~(\ref{sec:carbthree}).

\section{The chain coupling}\label{app:chcoupl}

Here we briefly discuss the proportional background solutions and the corresponding linearised equations for $N$ fields in a chain coupling. 
In the equations of motion~(\ref{eomchain}) we make the proportional ansatz,
\beqn
g^{i-1}_{\mu\nu}=c_i^2g^i_{\mu\nu}\,\,,
\qquad i=2,\hdots ,N\,.
\eeqn
The equations reduce to the following $N$ copies of Einstein's equations,
\beqn
\mathcal{G}_{\mu\nu}(g^i)+g^i_{\mu\nu}\Big(\frac{1}{\alpha_i^2}\Lambda(\beta_n^{(i)},c_i)
+ {c_{i+1}^{-2}}\tilde{\Lambda}(\beta_n^{(i+1)}, c_{i+1}, \alpha_i)\Big)
&=&0\,,\qquad i=2,\hdots N-1\,,\nn
\\
\mathcal{G}_{\mu\nu}(g^1)+g^1_{\mu\nu}\,{c_{2}^{-2}}\tilde{\Lambda}(\beta_{n}^{(2)},c_2, \alpha_1)&=&0\,,\nn\\
\mathcal{G}_{\mu\nu}(g^N)+g^N_{\mu\nu}\,\frac{1}{\alpha_N^2} \Lambda(\beta_{n}^{(N)}, c_N)&=&0\,,
\eeqn
where the cosmological constants are of the same functional form as in~(\ref{ccdef}).
The consistency conditions arising from the scale invariance of the Einstein tensor are,
\beqn
\Big(\frac{1}{\alpha_i^2}\Lambda(\beta_n^{(i)},c_i)+ {c_{i+1}^{-2}}\tilde{\Lambda}(\beta_n^{(i+1)}, c_{i+1}, \alpha_i)\Big)\prod_{k=2}^ic^{-2}_k
&=&{c_{2}^{-2}}\tilde{\Lambda}(\beta_{n}^{(2)},c_2, \alpha_1)\,,\qquad i=2,\hdots N-1\,,\nn\\
\frac{1}{\alpha_N^2} \Lambda(\beta_{n}^{(N)}, c_N)\prod_{k=3}^Nc^{-2}_k&=&\tilde{\Lambda}(\beta_{n}^{(2)},c_2, \alpha_1)\,.
\eeqn
The main complication with respect to the centre coupling arises because the proportionality constants appear in mixed and higher powers in these constraints.

Next we consider small perturbations, $\gmn^i=\bar{g}_{\mu\nu}^i+\delta\gmn^i$.
Expressing all quantities with respect to the background metric $\bar{g}^1_{\mu\nu}\equiv \bar{g}_{\mu\nu}$, we obtain the linearised equations around the above backgrounds as,
\begin{subequations}
\beqn
\alpha_i^2\bar{\mathcal{E}}^{~~\rho\sigma}_{\mu\nu}\delta g^i_{\rho\sigma} 
- m^2C_i\mathcal{A}_i\big( \delta S^i_{\mu\nu}-\delta S^i\bar{g}_{\mu\nu}\big)
+ \tfrac{m^2C_i\mathcal{A}_{i+1}}{c^2_{i+1}}\big( \delta S^{i+1}_{\mu\nu}-\delta S^{i+1}\bar{g}_{\mu\nu}\big)
=0\,,&~&\nn\\
i=2,\hdots, N-1\,, &~&
\\
\alpha_1^2\bar{\mathcal{E}}^{~~\rho\sigma}_{\mu\nu}\delta g^1_{\rho\sigma} 
+ \tfrac{m^2\mathcal{A}_{2}}{c_2^2}\big( \delta S^{2}_{\mu\nu}-\delta S^{2}\bar{g}_{\mu\nu}\big)
=0\,,&~&\\
\alpha_N^2\bar{\mathcal{E}}^{~~\rho\sigma}_{\mu\nu}\delta g^N_{\rho\sigma} 
- m^2C_N\mathcal{A}_N\big( \delta S^N_{\mu\nu}-\delta S^N\bar{g}_{\mu\nu}\big)
=0\,.&~&
\eeqn
\end{subequations}
Here the kinetic operator is the same as in (\ref{kinop}) and, for $i=1,\hdots, N$, we have defined the linear combinations of fluctuations,
\beqn
\delta S^i_{\mu\nu}\equiv\frac{1}{2c_i}\big(\delta  g^{i-1}_{\mu\nu}-c_i^2\delta\gmn^{i}\big)\,,
\eeqn
as well as the parameter combinations,
\beqn
\mathcal{A}_i\equiv  \beta^{(i)}_1+2c_i\beta^{(i)}_2+c_i^2\beta^{(i)}_3\,,
\qquad C_i\equiv\prod_{k=2}^ic^{-2}_k\,.
\eeqn
Deriving the mass spectrum (i.e.~diagonalising the mass matrix analogous to~(\ref{massmatrix})) is difficult in general. Nevertheless, we have verified that it always contains a massless spin-2 mode, as expected.


\end{document}